\documentclass[12pt,showpacs,showkeys,amsmath,amssymb]{revtex4}
\usepackage{amsmath,amsfonts,amsthm,amscd,amssymb,latexsym}
\usepackage{bm}
\usepackage{dcolumn}
\usepackage{graphicx}
\usepackage{epstopdf}
\usepackage{color}
\usepackage{epsf}
\usepackage{epsfig}
\usepackage{graphicx, epic, eepic, color}

\newcommand{\Hess}{\operatorname{Hess}}
\newcommand{\trace}{\operatorname{tr}}

\newcommand{\abs}[1]{\lvert#1\rvert}

\date{\today}

\begin{document}
\title{Cosmological Implications of Nonlocal Gravity}

\author{C. Chicone}
\email{chiconec@missouri.edu}
\affiliation{Department of Mathematics and Department of Physics and Astronomy, University of Missouri, Columbia,
Missouri 65211, USA}

\author{B. Mashhoon}
\email{mashhoonb@missouri.edu}
\affiliation{Department of Physics and Astronomy,
University of Missouri, Columbia, Missouri 65211, USA}

\begin{abstract}

We present extensions of the treatment contained in our recent paper on nonlocal Newtonian cosmology [C.~Chicone and B.~Mashhoon, J.\ Math.\ Phys.\  {\bf 57}, 072501 (2016)]. That is, the implications of the recent nonlocal generalization of Einstein's theory of gravitation are further investigated within the regime of Newtonian cosmology. In particular, we treat the nonlocal problem of structure formation for a spherically symmetric expanding dust model and show numerically that as the central density contrast grows, it tends to decrease slowly with radial distance as the universe expands. The nonlocal violation of Newton's shell theorem provides a physical interpretation of our numerical results. 

\end{abstract}



\maketitle

\section{Introduction}

Nonlocal gravity is a recent classical generalization of Einstein's theory of gravitation in which the gravitational field is local but satisfies field equations that are partial integro-differential equations. 
The nonlocal extension of general relativity (GR) has been realized via GR$_{||}$, the teleparallel equivalent of GR. GR$_{||}$ is the gauge theory of the Abelian group of spacetime translations. The formal analogy between GR$_{||}$ and electrodynamics originally led Friedrich W. Hehl to suggest that a nonlocal GR$_{||}$ could be developed in analogy with nonlocal electrodynamics. Friedrich's  idea has been constructive: the nonlocal part of the resulting nonlocal gravity theory may provide a natural explanation for ``dark matter"~\cite{NL1, NL2, NL3}. \emph{We dedicate this paper to Friedrich on the occasion of his eightieth birthday. }

Nonlocal gravity is a tetrad theory;  that is,  the gravitational potentials in nonlocal general relativity are given by the fundamental tetrad frame field $e_\mu{}^{\hat \alpha}(x)$ from which one obtains the spacetime metric via orthonormality, namely,  $g_{\mu \nu}(x)=e_\mu{}^{\hat \alpha}e_\nu{}^{\hat \beta}\eta_{\hat \alpha \hat \beta}$. Free test particles and null rays follow timelike and null geodesic of $g_{\mu \nu}$, respectively. In our convention,  the Minkowski metric tensor  $\eta_{{\hat{\alpha}} {\hat{\beta}}}$ is given by diag$(-1,1,1,1)$; moreover, Greek indices run from 0 to 3, while Latin indices run from 1 to 3. The hatted Greek indices ${\hat{\alpha}}$, ${\hat{\beta}}$, etc., refer to \emph{anholonomic} tetrad indices, while $\mu$, $\nu$, etc., refer to \emph{holonomic} spacetime indices. We use units such that $c=1$, unless  specified otherwise. The indices are raised and lowered by means of the metric tensors $g_{\mu \nu}(x)$ and $\eta_{{\hat{\alpha}} {\hat{\beta}}}$; furthermore, in order to change a holonomic index of a tensor into an anholonomic index or vice versa, we project the tensor on the fundamental tetrad field. The spacetime metric is compatible with the Levi-Civita connection as well as the Weitzenb\"ock connection.The fundamental tetrad frame field is globally teleparallel via the Weitzenb\"ock connection, which is curvature-free. Thus two distant vectors are considered parallel in nonlocal gravity if they have the same components with respect to their local fundamental tetrad frames. The curvature of the Levi-Civita connection and the torsion of the Weitzenb\"ock  connection are related and constitute complementary aspects of the gravitational field in nonlocal gravity. 

It is possible to express the field equation of GR in terms of the torsion of the Weitzenb\"ock  connection. The result is the gravitational field equation for the teleparallel equivalent of GR, namely, GR$_{||}$---see~\cite{Mashhoon:2014jna} and the references cited therein. Using the electromagnetic analogy, nonlocal gravity has been obtained from GR$_{||}$ by rendering it nonlocal by means of a ``constitutive" kernel. That is, the nonlocality is due to a causal constitutive kernel that is introduced into the theory in close analogy with the nonlocal electrodynamics of media. The kernel acts as the weight function for a certain average of the gravitational field over spacetime; indeed, in the absence of this nonocal contribution to the gravitational field equations, the theory reduces to Einstein's general relativity. 
Nonlocal gravity is thus history dependent and gravitational memory must therefore be taken into account. Memory fades in space and time and this circumstance must be reflected in the kernel of nonlocal gravity theory. It turns out that this simple spacetime memory of past events that is reflected in the nonlocal aspect of gravity simulates dark matter. That is, there is no dark matter in nonlocal gravity; instead, what appears as dark matter in astrophysics and cosmology is expected to be due to the nonlocal character of the gravitational interaction. 

Other than the trivial solution indicating the absence of a gravitational field in Minkowski spacetime, no exact solution of the field equation of nonlocal gravity is known at present; therefore, we must resort to the general linear approximation and its Newtonian limit~\cite{Mashhoon:2014jna}. To explore some of  the cosmological implications of nonlocal gravity, we have extended nonlocal gravity in the Newtonian regime to the cosmological domain~\cite{Chicone:2015sda}. Moreover, we have assumed  that nonlocal Newtonian cosmology is related to nonlocal gravity theory in much the same way as Newtonian cosmology is related to the standard homogeneous and isotropic cosmological models of GR~\cite{Chicone:2015sda}.  The purpose of this paper is to extend the treatment of Ref.~\cite{Chicone:2015sda} and discuss further some of the consequences of our cosmological model.  In particular, we study the formation of spherically symmetric structures within the framework of nonlocal Newtonian cosmology.

\section{Nonlocal Cosmological Model}

In the Newtonian regime of nonlocal gravity, Poisson's equation---namely, $\nabla^2 \Phi=4\pi G \,\rho$ for the gravitational potential $\Phi$ in terms of the density of matter $\rho$---is nonlocally modified such that 
\begin{equation}\label{CM1}
\nabla^2 \Phi=4\pi G \,(\rho+ \rho_{D})\,,
\end{equation}
where $\rho_D$ has the interpretation of the effective density of dark matter
\begin{equation}\label{CM2}
 \rho_D (t, \mathbf{x})=\int q(t, |\mathbf{x}-\mathbf{y}|)\, \rho(t, \mathbf{y})\,d^3y\,.
\end{equation}
Here, $q(t, r)$,  $r=|\mathbf{x}|$, is the \emph{reciprocal kernel} of nonlocal gravity in the Newtonian regime. Thus $\rho_D$ is given by the spatial convolution of the matter density with $q$. Within the framework of nonlocal gravity theory, $q(t, r)$ must satisfy certain requirements; moreover, it should account for the ``flat" rotation curves of spiral galaxies at the present epoch $(t=t_0)$~\cite{Mashhoon:2014jna}. Two possible functional forms for $q$ have been worked out explicitly, so that the reciprocal kernel is either
\begin{equation}\label{CM2a}
q_1(t_0, r)=\frac{1}{4\pi \lambda_0}~ \frac{1+\mu_0\,(a_0+r)}{r\,(a_0+r)}~e^{-\mu_0\, r}\,
\end{equation}
or
\begin{equation}\label{CM2b}
q_2(t_0, r)=\frac{1}{4\pi \lambda_0}~ \frac{1+\mu_0\,(a_0+r)}{(a_0+r)^2}~e^{-\mu_0\, r}\,.
\end{equation}
The kernel decays exponentially with increasing radial distance, since memory fades in space. Further comparison with observational data is expected to determine which reciprocal kernel is in better agreement with experiment.  In this connection, we should mention that at this stage of development of nonlocal gravity, the possibility that more complicated kernels may be required involving other parameters cannot be excluded. 

 At the present epoch $t=t_0$, $q(t_0, r)$ contains three length parameters $a_0$, $\lambda_0$ and  $\mu_0^{-1}$ and we assume that $a_0 < \lambda_0 < \mu_0^{-1}$. Here $a_0$ is the short-range parameter that remains to be determined, while $\lambda_0 \approx 3$ kpc and $\mu_0^{-1} \approx 17$ kpc are galactic lengths that have been tentatively determined from the rotation curves of nearby  spiral galaxies and the internal dynamics of nearby clusters of galaxies. It is clear from Eqs.~\eqref{CM2a} and~\eqref{CM2b} that $a_0$ and $\mu_0$ have to do with the shape of the kernel, while the Tohline--Kuhn parameter $\lambda_0$ determines its overall amplitude such that $q=0$ for $\lambda_0=\infty$. Moreover, it is possible to determine the radial gravitational force between two point particles that are a distance $r$ apart using Eqs.~\eqref{CM1}--\eqref{CM2b}, see Section V. The force of gravity turns out to be always attractive; in fact, for $0<r<a_0$, it is a linear superposition of Newton's inverse square force law and an attractive force that can be expressed as the sum of a series in powers of $r/a_0$. However, for the Tohline--Kuhn regime given by $a_0<r< \mu_0^{-1}$, the force is approximately a superposition of Newton's $1/r^2$ law and $1/(\lambda_0\,r)$, where the latter term is essentially responsible for the flat rotation curves of spiral galaxies~\cite{RaMa}. Finally, for $r>\mu_0^{-1}$, the force asymptotically approaches Newton's $1/r^2$ law again but with a gravitational constant $G\,(1+\tilde{\alpha}_0)$, which is Newton's constant  augmented by a factor  of $\tilde{\alpha}_0$ that depends on the three parameters $a_0$, $\lambda_0$ and $\mu_0^{-1}$; in fact, $\tilde{\alpha}_0\approx 10$. The force between point masses for $r\to \infty$ has a simple intuitive explanation in terms of the effective dark matter associated with a point mass. Further details about the force law are contained in Section V. We expect that the gravitational physics of the Solar System is affected by nonlocal gravity; indeed, using current Solar-System data regarding the perihelion precession of Saturn, a preliminary lower limit of order $10^{15}$ cm has been placed on $a_0$, see Ref.~\cite{ChMa}.

How can the functional form of $q(t, r)$ during past cosmological epochs be determined? In the absence of exact cosmological models~\cite{BiMa}, nonlocal Newtonian cosmology has been developed in close analogy with Newtonian cosmology~\cite{Chicone:2015sda}. That is, as in Newtonian cosmology~\cite{Pe, ZN, Mu}, we assume that after recombination nonlocal gravity is adequate for the description of the nonrelativistic motion of matter on subhorizon scales. In principle, the parameters of the reciprocal kernel  $q(t, r)$ could change with cosmic time; then, we would have $a(t)$, $\lambda(t)$ and $\mu^{-1}(t)$, such that  $a(t_0)=a_0$, etc. We note, based on inspection of Eqs.~\eqref{CM2a}--\eqref{CM2b}, that $a(t)$ and $\mu(t)$ would determine the shape of the kernel over short and long distances, respectively, while $\lambda(t)$ would determine its overall strength. Following recombination, once clumps of matter start to separate from the expanding background, they are expected to undergo internal gravitational collapse and eventually contribute to the formation of galaxies that are then very weakly affected by the gravitational tidal forces of the background expanding universe~\cite{Mashhoon:2007qm}. We therefore tentatively assume for the sake of simplicity that $a(t)=a_0$ and $\mu(t)=\mu_0$, since these parameters primarily affect the internal structure of self-gravitating galactic systems. On the other hand, the Tohline--Kuhn parameter $\lambda(t)$ determines the overall strength of the reciprocal kernel and this could be
 time-dependent.  Thus the main new idea is that the fading of memory in time implies that the nonlocal aspect of gravity must have been stronger in the past cosmological epochs; that is, the net strength of the gravitational interaction must decrease with cosmic time, since the \emph{effective amount of dark matter} decreases as the universe expands. To implement this idea, we assume that $\lambda(t)=B(t) \lambda_0$, where $B(t)$ monotonically increases with cosmic time and $B(t_0) = 1$. This means that we will henceforward assume
\begin{equation}\label{CM3}
 q(t, r) = \frac{q(t_0, r)}{B(t)}\,.
\end{equation}

\subsection{Nonlocal Newtonian Cosmology}

Newtonian cosmology reproduces essentially the same dynamics as the standard spatially homogeneous and isotropic Friedmann--Lema\^{\i}tre--Robertson--Walker (FLRW)  cosmological models of general relativity if pressure can be neglected. Working within the framework of nonlocal gravity in the Newtonian regime, we therefore imagine an infinite distribution of baryonic matter of density $\rho(t, \mathbf{x})$ and  zero pressure. The conservation of mass implies that the continuity equation should hold, namely, 
\begin{equation}\label{N1}
\partial_t \,\rho+\nabla\cdot (\rho \,\mathbf{v})=0\,,
\end{equation}
where $\rho\, \mathbf{v}$ is essentially the baryonic matter current. The acceleration of baryonic matter is due to the universal attraction of gravity; hence, Euler's equation of motion takes the form 
\begin{equation}\label{N2}
\frac{d\mathbf{v}}{dt}=\partial_t \,\mathbf{v} +(\mathbf{v}\cdot \nabla)\, \mathbf{v}=-\nabla\Phi\,,
\end{equation}
where $\Phi$ is the gravitational potential and satisfies the nonlocal Poisson equation 
\begin{equation}\label{N3}
\nabla^2 \Phi(t, \mathbf{x}) =4\pi G \,[\rho(t, \mathbf{x})+\int q(t, |\mathbf{x}-\mathbf{y}|)\, \rho(t, \mathbf{y})\,d^3y]\,.
\end{equation}

We are here interested in the solution of these equations for an infinite uniformly expanding  spatially homogeneous and isotropic perfect fluid medium with $\rho=\bar{\rho}(t)$ and vanishing pressure. We assume that this universe model expands in accordance with $\mathbf{x}=A(t)\,\boldsymbol{\xi}$, where $A(t)$ is the scale factor, $\boldsymbol{\xi}$ denotes the spatial position of the perfect fluid particle at the present epoch $t=t_0$ and  $\mathbf{x}$ denotes the spatial position of the particle at time $t$ such that $A(t_0)=1$. It follows that $\bar{\mathbf{v}}=d\mathbf{x}/dt = H(t)\, \mathbf{x}$, where $H(t)=\dot A /A$ is the Hubble parameter and an overdot denotes differentiation with respect to time $t$. Finally, let us note that in this uniform density case 
\begin{equation}\label{N4}
 \frac{\rho_D}{\rho} =  \int_{\mathbb{R}^3}q(t, |\mathbf{x}|)\,d^3x = \tilde{\alpha}(t)= \frac{\tilde{\alpha}(t_0)}{B}\,, \qquad \tilde{\alpha}(t_0)= \tilde{\alpha}_0 \approx 10\,.
\end{equation}

A comment is in order here regarding the effective dark matter fraction in our expanding universe model. Let $f_{DM}$ denote the ratio of the total mass of dark matter to the total mass of the baryonic matter in an astrophysical system. In the currently accepted model of standard cosmology, $f_{DM}$ for the universe is about 5 and independent of cosmic time. On the other hand, for nearby clusters of galaxies, for instance, $f_{DM}$ is about 10. This general circumstance leads to the problem of missing baryons in the standard model of cosmology~\cite{Shull:2011aa}.  Of course, the same problem of missing baryons could possibly exist in nonlocal gravity theory as well.  We emphasize that the nonlocal cosmological model under consideration here is a toy model. Nevertheless, the situation is quite different here with regards to the amount of effective dark matter; that is, in our nonlocal toy model, the effective dark matter fraction $f_{DM}$ for the universe monotonically  \emph{decreases} with cosmic time and is about 10 at the present epoch. 

With our assumptions, Eqs.~\eqref{N1}--\eqref{N3} have the solution
\begin{equation}\label{N5}
\bar{\rho}= A^{-3} \rho_0\,, \qquad  \bar{\mathbf{v}}=\dot A A^{-1} \mathbf{x}\,, \qquad
\bar{\Phi}=-\frac{1}{2} \ddot A A^{-1} r^2\,,
\end{equation}
where $\rho_0=\bar{\rho}(t_0)$ is the current density of baryonic matter in the universe, $\bar{\Phi}$ is determined up to an integration constant and the scale factor $A$ is a solution of the differential equation
\begin{equation}\label{N6}
A^{-1}\ddot A =-\frac{4\pi G \rho_0}{3} A^{-3}[1+\tilde{\alpha}(t)]\,.
\end{equation}
It remains to specify the monotonically increasing function $B(t)= \tilde{\alpha}_0/\tilde{\alpha}(t)$. Henceforth, we will assume that 
\begin{equation}\label{N7}
B(t) = A^{\varpi} (t)\,, \qquad  \varpi >0\,.
\end{equation}
Thus $B$ monotonically increases with cosmic time and $B(t_0)=1$. In this case, Eq.~\eqref{N6} can be integrated once and the result is
\begin{equation}\label{N8}
\frac{1}{2} \dot A^2= \frac{4\pi G\rho_0}{3} \left[\frac{1}{A}+\frac{\tilde{\alpha}_0}{(\varpi +1)\, A^{\varpi +1}}\right]+\bar{E}\,,
\end{equation}
where $\bar{E}$ is a constant of integration. To simplify matters, we set $\bar{E}=0$ in analogy with the critical case in Newtonian cosmology that corresponds to the spatially flat FLRW universe.  With $\bar{E} = 0$, Eq.~\eqref{N8} implies that  
\begin{equation}\label{N9}
H_0^2=\frac{8 \pi G \rho_0}{3}\, \left(1+\frac{\tilde{\alpha}_0}{\varpi +1}\right)\,,
\end{equation}
where $H_0=\dot A(t_0)$ is the Hubble constant. 

With the further assumption that $A(0)=0$ at the Big Bang, Eqs.~\eqref{N8} and~\eqref{N9} can be combined to find the age of the universe $t_0$ in units of the Hubble time $1/H_0$, namely,   
\begin{equation}\label{N10}
t_0\,H_0=\left(\frac{\varpi +1+\tilde{\alpha}_0}{\varpi +1}\right)^{1/2}\,\int_{0}^{1} \left[\frac{(\varpi +1)\,\tau^{\varpi +1}}{\tilde{\alpha}_0+(\varpi +1)\,\tau^{\varpi}}\right]^{1/2}\,d\tau\,,
\end{equation}
where only positive square roots are considered throughout. A graph of $t_0\,H_0$ versus $\varpi$ for 
$\tilde{\alpha}_0=11$ is presented in Figure 1. As $\varpi$ increases, the amount of  effective dark matter as well as the strength of the gravitational interaction increases monotonically toward the past epochs of the universe  and hence the age of the universe monotonically decreases; this behavior is evident in Figure 1. 

\begin{figure}
\begin{center}
\includegraphics[width=12 cm]{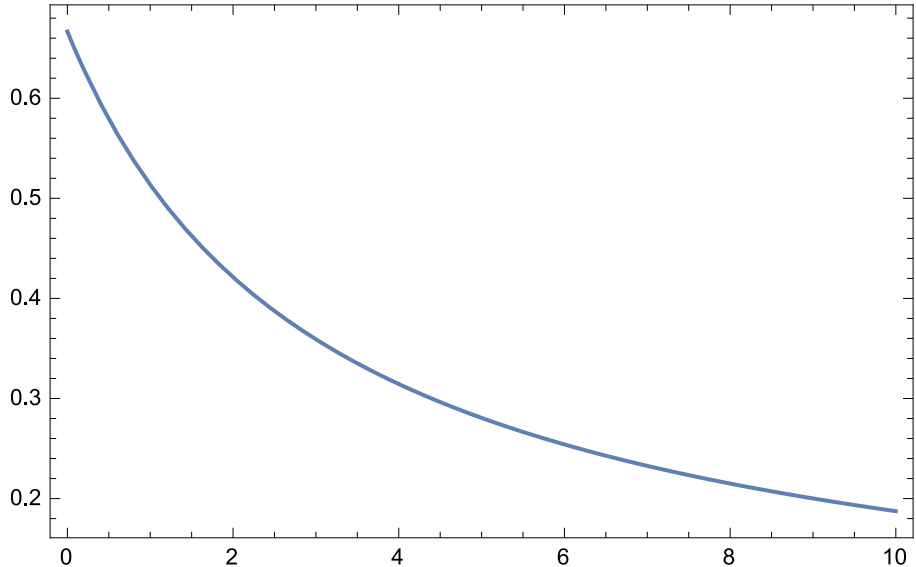}
\end{center}
\caption{Plot of $t_0/(1/H_0)$ versus $\varpi$ with $\tilde{\alpha}_0=11$. The age of the universe in units of $1/H_0$ monotonically decreases with increasing $\varpi$, as expected.}
\label{fig1} 
\end{figure}

To recover standard Newtonian cosmology, we must formally set $\tilde{\alpha}_0=0$. Moreover, in Eq.~\eqref{N6}, $\tilde{\alpha}(t)$, with $B(t)$ given by Eq.~\eqref{N7}, monotonically decreases with time and goes to zero as $t\to \infty$; therefore, the solution of Eq.~\eqref{N8} asymptotically approaches the Newtonian analog of the Einstein--de Sitter model, $A(t) \propto t^{2/3}$, as $t\to \infty$.

\subsection{Jeans Instability}

For $\varpi=1$, a detailed treatment of the resulting cosmological model is contained in Ref.~\cite{Chicone:2015sda}, where it is shown, via a Jeans stability analysis, that the expanding perfect-fluid medium is linearly unstable to  structure formation for adiabatic perturbations on scales that are much larger than the Jeans length; moreover, when gravitational instability takes over, the fluid pressure may be neglected~\cite{Pe, ZN, Mu}.  The linear stability analysis can be straightforwardly extended to the general case of $\varpi >0$. The problem of large scale structure formation in the universe was considered in Ref.~\cite{Chicone:2015sda} following the approach originally developed by Zeldovich~\cite{YaBZ}, which is crucial for the understanding of the cosmic web~\cite{Mu, YaBZS, GSS, Hidd}.  The dependence of structure formation upon $\varpi$ is an interesting problem, to which we now turn.

\section{Structure Formation}

To investigate the nonlinear instability of our model, it is useful to express the continuity equation as 
\begin{equation}\label{Z1}
[\partial_t +(\mathbf{v}\cdot \nabla)]\, \rho + \rho\, \nabla\cdot \mathbf{v}=0\,
\end{equation}
and to combine Eqs.~\eqref{N2} and~\eqref{N3} by eliminating $\Phi$ between them, namely,
\begin{equation}\label{Z2}
\nabla\cdot [\partial_t \,\mathbf{v} +(\mathbf{v}\cdot \nabla)\, \mathbf{v}]=-4\pi G \,(\rho+ \rho_{D})\,.
\end{equation}

Following Zel'dovich~\cite{YaBZ}, we wish to investigate the nonlinear gravitational instability of Eqs.~\eqref{Z1}--\eqref{Z2} using  Lagrangian variables. The Lagrangian variables form a spatial coordinate system such that the Lagrangian coordinates $\boldsymbol{\xi}$ of a fluid particle uniquely identify the particle and are constants along its trajectory~\cite{Mu}. 

To transform Eqs.~\eqref{Z1}--\eqref{Z2} to Lagrangian coordinates $\boldsymbol{\xi}$, we assume that   the position of a fluid particle at a time $t$ is given by $\mathbf{x}=\mathbf{x}(t, \boldsymbol{\xi})$. For example, coordinates $\boldsymbol{\xi}$ could specify the spatial positions of fluid particles  at some  initial epoch $t_{in}>0$ after the Big Bang, i.e., $\mathbf{x}(t_{in})= \boldsymbol{\xi}$. Lagrangian variables are admissible so long as the fluid trajectories with different Lagrangian coordinates do not meet in space; otherwise, the Lagrangian coordinate system breaks down with the attendant formation of caustics. In relativity theory, the comoving coordinate system is the natural generalization of the Lagrangian coordinate system of Newtonian mechanics. 

The fluid velocity can be written in Lagrangian variables as 
\begin{equation}\label{Z3}
\mathbf{v}=\frac{d\mathbf{x}}{dt}=\left . \frac{\partial \mathbf{x}}{\partial t}(t, \boldsymbol{\xi})\right |_{\boldsymbol{\xi}}\,.
\end{equation}
In general,
\begin{equation}\label{Z4}
\left . \frac{\partial~}{\partial t}\right|_{\mathbf{x}} + \mathbf{v}\cdot \nabla_{\mathbf{x}}= \left . \frac{\partial~}{\partial t}\right |_{\boldsymbol{\xi}}\,.
\end{equation} 
To transform the continuity Eq.~\eqref{Z1} to Lagrangian variables, we note that 
\begin{equation}\label{Z5}
\nabla_j\,v^i= \frac{\partial \xi^k}{\partial x^j}\,\frac{\partial v^i(t, \boldsymbol{\xi})}{\partial \xi^k}= \frac{\partial \xi^k}{\partial x^j}\,\frac{\partial}{\partial t}\left(\frac{\partial x^i(t, \boldsymbol{\xi})}{\partial \xi^k}\right)\,;
\end{equation}
therefore, it proves useful to introduce a matrix $\Xi$ and its inverse $\Xi^{-1}$ with components
\begin{equation}\label{Z6}
\Xi^i{}_j:= \frac{\partial x^i}{\partial \xi^j}\,, \qquad  (\Xi^{-1})^i{}_j:= \frac{\partial \xi^i}{\partial x^j}\,,
\end{equation}
such that Eq.~\eqref{Z1} can be written as 
\begin{equation}\label{Z7}
\frac{\partial \varrho(t, \boldsymbol{\xi})}{\partial t} + \varrho \, {\rm tr} \left(\frac{\partial \Xi}{\partial t}\,\Xi^{-1}\right)=0\,,\end{equation}
where $\varrho$ is the density of baryons in Lagrangian coordinates, namely, 
\begin{equation}\label{Z8}
\rho(t, \mathbf{x}(t, \boldsymbol{\xi})):=\varrho(t, \boldsymbol{\xi})\,.
\end{equation}
Next, we define
\begin{equation}\label{Z9}
\mathbb{J}:=\det \Xi\,
\end{equation}
and recall the general mathematical result  
\begin{equation}\label{Z10}
\delta \mathbb{J}= \mathbb{J}\, (\delta \Xi)^i{}_j\, (\Xi^{-1})^j{}_i\,.
\end{equation}
It then follows that
\begin{equation}\label{Z11}
\frac{\partial \mathbb{J}}{\partial t}= \mathbb{J}\, {\rm tr} \left(\frac{\partial \Xi}{\partial t}\,\Xi^{-1}\right)\,
\end{equation}
and the continuity Eq.~\eqref{Z1} can be expressed as
\begin{equation}\label{Z12}
\frac{\partial (\varrho \, \mathbb{J})}{\partial t}=0\,.
\end{equation}

We now turn to Eq.~\eqref{Z2} and write the left-hand side of this equation as 
\begin{equation}\label{Z13}
\nabla\cdot [\partial_t \,\mathbf{v} +(\mathbf{v}\cdot \nabla)\, \mathbf{v}]= \left(\frac{\partial~}{\partial t}+ \mathbf{v}\cdot \nabla \right)\,\nabla \cdot \mathbf{v} + (\nabla_i\,v^j)(\nabla_j\,v^i)\,.
\end{equation}
The divergence of the fluid velocity in Lagrangian variables is $\partial \ln{\mathbb{J}}/\partial t$, which can be employed to transform Eq.~\eqref{Z2} to 
\begin{equation}\label{Z14}
\frac{\partial^2}{\partial t^2} \ln \mathbb{J}+{\rm tr} \left[\left(\frac{\partial \Xi}{\partial t}\, \Xi^{-1}\right)^2\right]=-4\pi G(\varrho+\varrho_D)\,.
\end{equation}
Here,  $\varrho_D$ is the density of the effective dark matter expressed in Lagrangian coordinates. It proves advantageous to study the solutions of the Lagrangian Eqs.~\eqref{Z12} and~\eqref{Z14} instead of the Eulerian Eqs.~\eqref{Z1}--\eqref{Z2}. 

It is interesting to illustrate the use of Lagrangian coordinates by solving Eqs.~\eqref{Z12} and~\eqref{Z14} for the spatially homogeneous and isotropic expanding background spacetime. In this case, let us assume
\begin{equation}\label{Z15}
\mathbf{x}(t, \boldsymbol{\xi})=b(t)\, \boldsymbol{\xi}\,,
\end{equation}  
where $b(t_{in})=1$. Then, $\Xi=b(t)$\,diag$(1,1,1)$, $\mathbb{J}=b^3(t)$ and Eq.~\eqref{Z12} implies that $\varrho\,\mathbb{J} = \varrho(t_{in}, \boldsymbol{\xi})$. However,  it follows from the homogeneity of the background that $\varrho$ is independent of $\boldsymbol{\xi}$. Furthermore, we find from Eq.~\eqref{Z14} that $3\,\ddot{b}/b = -4 \pi G (1+\varrho_D/\varrho)\,\varrho$. We recall that in this homogeneous case, $\varrho_D/\varrho=\tilde{\alpha}(t)$ and $\varrho(t)=\varrho(t_{in})/b^3(t)$.  In this way, we recover Eqs.~\eqref{N5}--\eqref{N6} with $b(t)=A(t)/A(t_{in})$, $\varrho=\bar{\rho}$ and $\rho_0=\varrho(t_{in})\,A^3(t_{in})$.

\subsection{Zel'dovich Ansatz}

To go beyond  the homogeneous solution~\eqref{Z15}, we assume a Lagrangian fluid flow of the form
 \begin{equation}\label{Z16}
 \mathbf{x}(t, \boldsymbol{\xi})=\mathbb{A}(t)\,[\boldsymbol{\xi}-\mathbf{F}(t, \boldsymbol{\xi})]\,,
 \end{equation}
where 
 \begin{equation}\label{Z17}
 \mathbf{F}(t, \boldsymbol{\xi}):= (F^1(t, \xi^1), 0, 0)\,.
 \end{equation}
In this case, we have
\begin{equation}\label{Z18}
\Xi=\mathbb{A}(t)\, {\rm diag}(1-\Psi, 1, 1)\,, \qquad \mathbb{J}=\mathbb{A}^3\,(1-\Psi)\,, \qquad \Psi := \frac{\partial F^1}{\partial \xi^1}\,.
\end{equation}
It follows from the continuity Eq.~\eqref{Z12} that
\begin{equation}\label{Z19}
\varrho=\frac{\varrho_0(\boldsymbol{\xi})}{\mathbb{A}^3\,(1-\Psi)}\,,
\end{equation}
where $\varrho_0(\boldsymbol{\xi})>0$ is simply a function of the Lagrangian coordinates that are constants of the motion along the trajectory of a fluid particle. Next, we substitute  the 
Zel'dovich ansatz~\eqref{Z16} into Eq.~\eqref{Z14} and find 
 \begin{equation}\label{Z20}
- \,3\,\frac{\ddot{\mathbb{A}}}{\mathbb{A}} 
 +2\, \frac{\dot{\mathbb{A}}}{\mathbb{A}}\, \frac{\Psi_{t}} {1-\Psi}+ \frac{\Psi_{tt}}{1-\Psi}= 4\pi G\,\Big(\frac{\varrho_0(\boldsymbol{\xi})}{\mathbb{A}^3(1-\Psi)} +\varrho_D\Big)\,.
 \end{equation}
 
In accordance  with the original Zel'dovich solution~\cite{Mu}, we assume
\begin{equation}\label{Z21}
\mathbb{A}=A\,, \qquad \varrho_0(\boldsymbol{\xi}) = \rho_0\,,
\end{equation}
where $A(t)$ is the scale factor for  the expanding spatially homogeneous and isotropic background and $\rho_0$ is the uniform background baryonic density at the present epoch.  Substituting  
Eqs.~\eqref{N6} and~\eqref{N7}  in  Eq.~\eqref{Z20}, we find
 \begin{equation}\label{Z22}
 \frac{\Psi_{tt}}{1-\Psi}
 +2\, \frac{\dot{A}}{A}\,\frac{\Psi_{t}} {1-\Psi}+\frac{4\pi G\rho_0 }{A^3}(1+\frac{\tilde{\alpha}_0}{A^\varpi})=4\pi G\, \Big(\frac{\rho_0}{A^3(1-\Psi)} +\varrho_D\Big)\,.
 \end{equation}
 
The density of baryons in this nonlocal Zel'dovich model is different from the homogeneous background model by a factor of  $(1-\Psi)^{-1}$.  In fact, the density contrast is given by $\Psi/(1-\Psi)$, and one expects that for a density contrast of five, say, that is sufficiently large compared to unity, the over-dense region separates from the background and collapses under its own gravity~\cite{Mu}. That is, for $\Psi=0$, we recover  the homogeneous background, while for  $\Psi=1$ we have a caustic singularity that indicates the breakdown of the Lagrangian coordinate system. On the other hand, when $\Psi$ is sufficiently close to unity, the baryonic density can be high enough that a clump of matter would collapse under its own gravity and would therefore separate from the background thus forming cosmic structure. It is possible to write Eq.~\eqref{Z22} in the form 
 \begin{equation}\label{Z23}
 \Psi_{tt}+2\,\frac{\dot{A}}{A}\, \Psi_{t}-\frac{4\pi G\rho_0}{A^3}\,(1+\frac{\tilde{\alpha}_0}{A^\varpi})\, \Psi =\mathcal{N}\,,
 \end{equation}
where   $\mathcal{N}$  is defined by
 \begin{equation}\label{Z24}
\mathcal{N} := \frac{4\pi G\rho_0\,\tilde{\alpha}_0}{A^{3+\varpi}}\Big[\frac{A^{3+\varpi}}{\rho_0\,\tilde{\alpha}_0}\,(1-\Psi)\,\varrho_D-1\Big]\,.
 \end{equation} 
 
Let us briefly digress here and mention that in the \emph{local} gravitation theory with $\tilde{\alpha}_0=0$, we have $\mathcal{N}=0$ in Eq.~\eqref{Z23}  and we obtain in this way the original \emph{Zel'dovich solution}~\cite{Mu, YaBZ}. That is, Eq.~\eqref{Z23} implies that $\Psi$ is in this case a linear combination of $A$ and $A^{-3/2}$, which correspond, respectively, to the growing and decaying modes of the linearized theory~\cite{Mu}. The result of the linearized theory is thus naturally extended to the nonlinear regime in the Zel'dovich solution. The exact one-dimensional solution for the growing mode was then generalized by Zel'dovich to obtain an approximate solution of the general nonlinear three-dimensional  equations~\cite{YaBZ}. This \emph{Zel'dovich approximation} plays a significant role in the theories of the large scale structure of the universe and in the interpretation of the cosmic web~\cite{YaBZS, GSS, Hidd}.
 
It is important to note that   $\mathcal{N}$ vanishes for $\varrho_D=(\tilde{\alpha}_0/A^\varpi)\, \varrho$, which is exactly the same as Eq.~\eqref{N4} for a uniform density configuration. In fact, as discussed in detail in Ref.~\cite{Chicone:2015sda}, this condition is approximately satisfied and $\mathcal{N}\approx 0$ for perturbations in baryonic density over scales that are much larger than $\mu_0^{-1}$. Therefore, on such large scales that persist over time, the dimensionless density contrast function $\Psi$ should satisfy
 \begin{equation}\label{Z25}
 \Psi_{tt}+2\,\frac{\dot{A}}{A}\, \Psi_{t}-\frac{4\pi G\rho_0}{A^3}\,(1+\frac{\tilde{\alpha}_0}{A^\varpi})\, \Psi =0\,.
 \end{equation} 
 For $\tilde{\alpha}_0\ne 0$, we seek a solution of Eq.~\eqref{Z25} that is of the form
 \begin{equation}\label{Z26}
 \Psi = A^{\bar{\sigma}}\, \chi(\nu)\,, \qquad \nu=-\frac{\varpi +1}{\tilde{\alpha}_0}\, A^{\varpi}\,. 
\end{equation}  
Substituting Eq.~\eqref{Z26} in Eq.~\eqref{Z25}, we find 
 \begin{equation}\label{Z27}
 \nu^2\,(1-\nu)\,\chi_{\nu \nu} +\nu \left[\frac{1}{2}+\frac{4\bar{\sigma} +1}{2\,\varpi} - \left(1+ \frac{4\bar{\sigma} +1}{2\,\varpi}\right)\,\nu\right] \,\chi_\nu + \left(\frac{\mathbb{Q}}{2\,\varpi^2} - \frac{\bar{\sigma} +3}{2\,\varpi}\,\nu \right)\,\chi = 0\,, 
\end{equation}  
where
\begin{equation}\label{Z28}
 \mathbb{Q} = 2\,\bar{\sigma}^2 - (\varpi-1)\,\bar{\sigma} -3\,(\varpi +1)\,.
\end{equation} 

Let us determine $\bar{\sigma}$ by setting $\mathbb{Q}=0$; that is,  $\bar{\sigma}=\sigma_{\pm}$ given by
\begin{equation}\label{Z29}
\sigma_{\pm}(\varpi) = \frac{1}{4}\,[\varpi-1\pm(\varpi^2+22\,\varpi+25)^{1/2}]\,,
\end{equation}
where $\sigma_{+} >0$ and $\sigma_{-} <0$. We recall that $\varpi >0$ by assumption; on the other hand, the limiting case $\varpi =0$ in Eq.~\eqref{Z29} leads to $\sigma_{+} =1$ and $\sigma_{-} =-3/2$, just as in the Zel'dovich solution. With $\mathbb{Q}=0$, Eq.~\eqref{Z27} reduces to the hypergeometric equation
 \begin{equation}\label{Z30}
 \nu\,(1-\nu)\,\chi_{\nu \nu} +[\,\hat{c} -  (\hat{a} + \hat{b} +1)\,\nu\,] \,\chi_\nu -\hat{a}\,\hat{b} \,\chi= 0\,,
 \end{equation} 
where
\begin{equation}\label{Z31}
 \hat{a}=\frac{2\bar{\sigma} +3}{2\,\varpi}\,, \qquad \hat{b} = \frac{\bar{\sigma} -1}{\varpi}\,, \qquad \hat{c}= \frac{1}{2}+\frac{4\bar{\sigma} +1}{2\,\varpi}\,.
\end{equation} 
We note that  $\hat{c} = \frac{1}{2} + \hat{a} + \hat{b}$. The solution of Eq.~\eqref{Z30} that is regular in 
$A$ is in the form of the hypergeometric series  $F(\hat{a}, \hat{b}; \hat{c}; \nu)$, which converges for $|\nu| \le 1$~\cite{A+S}. Therefore, the general solution for $\Psi$ takes the form
\begin{equation}\label{Z32}
 \mathcal{S}_{\bar{\sigma}}(A) = A^{\bar{\sigma}}\,F\left(\hat{a}, \hat{b}; \hat{c}; -\frac{\varpi +1}{\tilde{\alpha}_0}\, A^{\varpi}\right)\,.
\end{equation} 
Thus a fundamental set of solutions of the linear second-order Eq.~\eqref{Z25} is given by the growing mode $\mathcal{S}_{\sigma_{+}}$ and the decaying mode $\mathcal{S}_{\sigma_{-}}$. The general solution of Eq.~\eqref{Z25} is thus qualitatively similar to the Zel'dovich solution. 

In the standard cosmological models, it is believed that at the epoch of decoupling, corresponding to 
$A \approx 10^{-3}$, small amplitude inhomogeneities of order $\approx 10^{-5}$ existed that have then grown via gravitational instability to produce the observed structure in the universe. In this scenario, the nature of processes that have produced galaxies, clusters of galaxies and eventually the cosmic web  is the subject of numerous investigations; in particular, it is generally  believed that \emph{dark matter} of unknown origin has played a crucial role in this development~\cite{Pe, ZN, Mu, GSS, Hidd}. In nonlocal gravity, there is effective dark matter due to gravitational memory. Thus an important question is whether the effective dark matter of nonlocal gravity can ensure the development of the observed structure in the universe. In the absence of exact (cosmological) solutions of nonlocal gravity, we have resorted to the  present toy model of nonlocal Newtonian cosmology. In this model, as $\Psi \to 1$, the density of baryons $\rho_0/[A^3\,(1-\Psi)]$ approaches infinity, which means that at some epoch before the singularity, the density contrast is so large that the perturbation separates from the background and collapses under its own gravitational attraction, thus leading in the end to the formation of structure in the universe. 
 In Fig. 2, we concentrate on the growing mode and for different values of $\varpi$ plot 
\begin{equation}\label{Z33}
 10^{-5}\,\frac{\mathcal{S}_{\sigma_{+}}(A)}{\mathcal{S}_{\sigma_{+}}(10^{-3})}\,
\end{equation} 
versus $A$. Here,  it is assumed that at the epoch of recombination corresponding to the scale factor of   $A \approx 10^{-3}$, 
the density contrast is $\approx 10^{-5}$. It is demonstrated in Fig. 2 that $\Psi$ for the growing mode approaches unity well  before the present era ($A=1$) for $\varpi \gtrsim 1$. As $\varpi$ increase, structure formation occurs earlier in the history of the universe. Thus structure formation is theoretically possible in this toy model if $\varpi$ is large enough, namely,  $\varpi \gtrsim 1$.

\begin{figure}
\begin{center}
\includegraphics[width=12 cm]{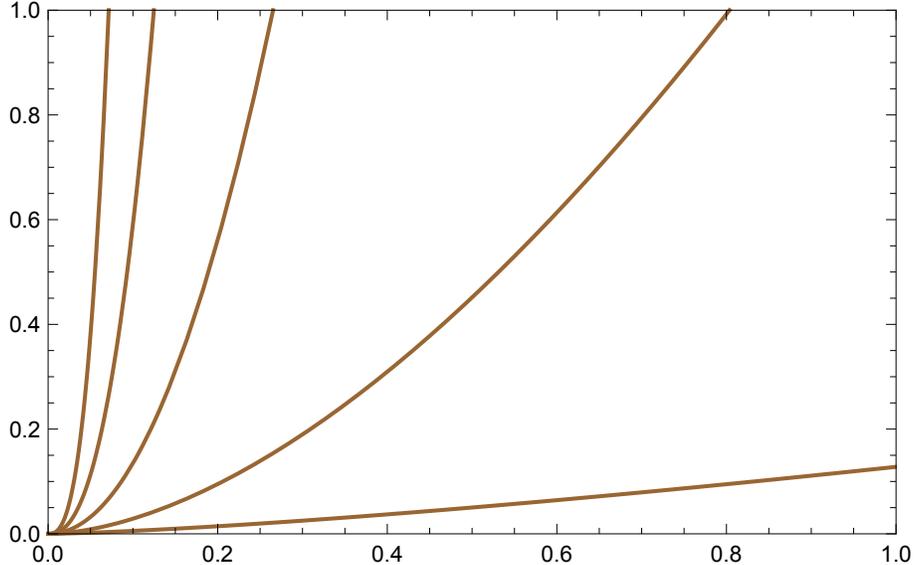}
\end{center}
\caption{Plot of $A \mapsto 10^{-5}\,\mathcal{S}_{\sigma_{+}}(A)/\mathcal{S}_{\sigma_{+}}(10^{-3})$ for  values of the parameter $\varpi= 0.5, 1, 1.5, 2, 2.5$, increasing from the right graph to the left.}
\label{fig2} 
\end{figure}

\subsection{Nonlocal Newtonian LTB Models}

It is interesting to consider cosmological dust models with spherically symmetric inhomogeneities~\cite{Mu}. Indeed, it is possible to extend the one-dimensional spatial inhomogeneity of the Zel'dovich solution to the radial inhomogeneity of the Newtonian analogs of LTB models.  The LTB models are spherically symmetric inhomogeneous dust solutions of general relativity that were first discovered by Lema\^{\i}tre~\cite{GLe}. They have been subsequently studied by Tolman~\cite{Tol}, Bondi~\cite{HBo} and many others; see, for example, Ref.~\cite{PK} for a detailed discussion.

We start with 
\begin{equation}\label{L1}
 x^i=R(t, \ell)\,\ell^i\,,
\end{equation} 
where $\boldsymbol{\ell} = (\ell^1, \ell^2, \ell^3)$ are the Lagrangian coordinates of a fluid particle in this case and $\ell=|\boldsymbol{\ell}|$ is the radial Lagrangian coordinate. Therefore, 
\begin{equation}\label{L2}
\Xi^i{}_j = R\, \delta^i_j + \frac{1}{\ell} R'\, \ell^i\,\ell_j\,,  
\end{equation}
where a prime denotes partial differentiation with respect to $\ell$, i.e. $R'=\partial R/\partial \ell$. Next, 
\begin{equation}\label{L3}
\mathbb{J} =\det \Xi=R^2\,S'\,,
\end{equation}
where
\begin{equation}\label{L4}
S(t, \ell):=\ell \, R(t, \ell)\,.
\end{equation}

It is possible to show that 
\begin{equation}\label{L5}
 (\Xi^{-1})^i{}_j =\frac{1}{R}\, \delta^i_j - \frac{R'}{S\,S'} \, \ell^i\,\ell_j\,.
\end{equation}
From
\begin{equation}\label{L6}
\left(\frac{\partial \Xi}{\partial t}\right)^i{}_j = \dot{R}\, \delta^i_j + \frac{1}{\ell} \dot{R}'\, \ell^i\,\ell_j\,,  
\end{equation}
where a dot denotes partial differentiation with respect to $t$, and Eq.~\eqref{L5}, we find 
\begin{equation}\label{L7}
\left(\frac{\partial \Xi}{\partial t}\,\Xi^{-1}\right)^i{}_j= \frac{\dot{S}}{S}\,\delta^i_j + \frac{1}{\ell^2}\,\left(\frac{\dot{S}'}{S'} - \frac{\dot{S}}{S}\right)\, \ell^i\,\ell_j\,.
\end{equation}
It follows that 
\begin{equation}\label{L8}
{\rm tr} \left[\left(\frac{\partial \Xi}{\partial t}\, \Xi^{-1}\right)^2\right]= 2\, \left(\frac{\dot{S}}{S}\right)^2 +\left(\frac{\dot{S}'}{S'}\right)^2\,.
\end{equation}
On the other hand, 
\begin{equation}\label{L9}
\frac{\partial^2}{\partial t^2} \ln \mathbb{J}= 2\,\left( \frac{\ddot{S}}{S} -  \frac{\dot{S}^2}{S^2}\right) +  \frac{\ddot{S}'}{S'}\,.
\end{equation}
Putting Eqs.~\eqref{L8} and~\eqref{L9} together, we have 
\begin{equation}\label{L10}
\frac{\partial^2}{\partial t^2} \ln \mathbb{J}+{\rm tr} \left[\left(\frac{\partial \Xi}{\partial t}\, \Xi^{-1}\right)^2\right]= 2\,  \frac{\ddot{S}}{S}+ \frac{\ddot{S}'}{S'} = \frac{(S^2\,\ddot{S})'}{S^2\,S'}\,.
\end{equation}

It is a consequence of the continuity Eq.~\eqref{Z12} that in this spherically symmetric configuration
\begin{equation}\label{L11}
\varrho (t, \ell) = \frac{\varrho_0(\ell)}{R^2\,S'}\,,
\end{equation}
where the density $\varrho_0(\ell) >0$ is an integration function,  $R\ge 0$  and $S'>0$. The Lagrangian coordinate system breaks down and caustic singularities occur for $S' = 0$. 

Equation~\eqref{Z14} can now be written as 
\begin{equation}\label{L12}
(S^2\,\ddot{S})' =-4\pi G\,[\ell^2\,\varrho_0(\ell) + S^2\,S'\, \varrho_D]\,,
\end{equation}
where
\begin{equation}\label{L13}
 \varrho_D (t, \ell)=\int q(t, |S(t, \ell)\,\widehat{\boldsymbol{\ell}}-S(t, \zeta)\,\widehat{\boldsymbol{\zeta}}|)\, \varrho_0(\zeta)\,d^3\zeta\,.
\end{equation}
Here, $\widehat{\boldsymbol{\ell}}=\boldsymbol{\ell}/\ell$ and 
$\widehat{\boldsymbol{\zeta}}=\boldsymbol{\zeta}/\zeta$ are unit vectors. 
Equations~\eqref{L11}--\eqref{L13} characterize the nonlocal Newtonian LTB models under consideration here. 

It is possible to recover the spatially homogeneous and isotropic cosmological background~\eqref{N5} from Eqs.~\eqref{L11}--\eqref{L13}. In this case, $\varrho_0(\ell) = \rho_0$ and $R(t, \ell)=A(t)$; moreover, $\varrho_D/ \varrho =\tilde{\alpha}_0/A^{\varpi}$ as a consequence of Eq.~\eqref{N4}. It then follows from Eq.~\eqref{L12} that
\begin{equation}\label{L13a}
A^2\,\ddot A =-\frac{4\pi G \rho_0}{3} \,\left(1+\frac{\tilde{\alpha}_0}{A^\varpi}\right)\,,
\end{equation}
in agreement with Eq.~\eqref{N6}. 

It appears that extensive numerical work is needed to solve the integro-differential Eq.~\eqref{L12} for $S=\ell\,R(t, \ell)$. For spherically symmetric perturbations over scales much larger than $\mu_0^{-1}$, we may assume, as before, that  $\varrho_D \approx (\tilde{\alpha}_0/A^{\varpi})\,\varrho$, in which case Eq.~\eqref{L12} reduces to
\begin{equation}\label{L14}
(S^2\,\ddot{S})' =-4\pi G\,\ell^2\,\varrho_0(\ell)\,\left(1+\frac{\tilde{\alpha}_0}{A^{\varpi}}\right)\,,
\end{equation}
where $A(t)$ is the scale factor. It is now possible to integrate Eq.~\eqref{L14} once to get
\begin{equation}\label{L15}
R^2\,\ddot{R} =-\frac{4\pi G}{3}\,\left(1+\frac{\tilde{\alpha}_0}{A^{\varpi}}\right)\,\bar{\varrho}(\ell)\,,
\end{equation}
where 
\begin{equation}\label{L16}
\bar{\varrho}(\ell) :=\frac{3}{\ell^3}\,\int_0^\ell \varrho_0(x)\,x^2\,dx\,.
\end{equation}
It does not appear possible to solve Eq.~\eqref{L15} analytically by replacing $t$ with $A(t)$ as the independent variable; in any case, we do not know a general explicit solution of Eq.~\eqref{L15}.  Furthermore, the general nonlocal problem in this Lagrangian formulation seems intractable. 

In the absence of nonlocality, i.e. when we formally let  $\tilde{\alpha}_0=0$, we recover the standard solutions of the local LTB models~\cite{Mu, PK}. That is, Eq.~\eqref{L15} with  
$\tilde{\alpha}_0=0$ can be integrated once with the result that 
\begin{equation}\label{L17}
\frac{1}{2}\,{\dot{R}}^2 = \frac{4\pi G\,\bar{\varrho}(\ell)}{3}\,\frac{1}{R} + E(\ell)\,,
\end{equation}
where the energy function $E(\ell)$ is an arbitrary function of integration. Then, for $E(\ell)<0$, we have 
\begin{equation}\label{L18}
R(t, \ell) =- \frac{2\pi G\,\bar{\varrho}}{3\,E}\,(1-\cos \eta)\,
\end{equation}
and
\begin{equation}\label{L19}
t-t_B(\ell) = \frac{4\pi G\,\bar{\varrho}}{3\,(-2E)^{3/2}}\,(\eta-\sin \eta)\,.
\end{equation}
For $E(\ell)=0$, we find 
\begin{equation}\label{L20}
R(t, \ell) = (6\pi G\,\bar{\varrho})^{1/3}\, [t-t_B(\ell)]^{2/3}\,.
\end{equation}
Finally, for $E(\ell)>0$, we have 
\begin{equation}\label{L21}
R(t, \ell) =\frac{2\pi G\,\bar{\varrho}}{3\,E}\,(\cosh \eta -1)\,
\end{equation}
and
\begin{equation}\label{L22}
t-t_B(\ell) = \frac{4\pi G\,\bar{\varrho}}{3\,(2E)^{3/2}}\,(\sinh \eta - \eta)\,.
\end{equation}
In these solutions, $t_B(\ell)$ is the bang-time function, since the time of the Big Bang singularity in general depends upon the radial position $\ell$.  The Big Bang singularity occurs at $R(t_B(\ell), \ell) =0$, where $\eta=0$. The local Newtonian LTB models should be compared and contrasted with the corresponding general relativistic models that are described in detail in Ref.~\cite{PK}. 

 We will return to the Eulerian formulation of the nonlocal problem of spherically symmetric structure formation in Section VI, since known methods of computational fluid dynamics can be employed in the numerical investigation of this problem within the Eulerian framework. 
 
Finally, it is interesting to note that the caustics that play an essential role in the Zel'dovich approach are indeed the shell-crossing singularities that naturally appear as a consequence of gravitational instability of fluid masses. This point is elucidated within the framework of nonlocal gravity in the next section.

\section{Raychaudhuri analog for Euler--Poisson dust model}

In nonlocal gravity, the geometry of a congruence of timelike (or null) curves can be developed using the Levi-Civita connection just as in general relativity~\cite{HE}. In the corresponding Raychaudhuri equation~\cite{Ray}, however,  the term $R_{\mu \nu}\, K^\mu K^\nu$ for a timelike vector $K$ cannot be treated as in general relativity, since the generalization of Einstein's field equation 
\begin{equation}\label{R1}
 R_{\mu \nu}-\frac{1}{2}\,R\, g_{\mu \nu} +\Lambda\, g_{\mu \nu}=8\pi G\, T_{\mu \nu}\,
\end{equation}
in nonlocal gravity is~\cite{BiMa}
\begin{equation}\label{R2}
 R_{\mu \nu}-\frac{1}{2}\,R\, g_{\mu \nu} +\Lambda\, g_{\mu \nu}=8\pi G\, (T_{\mu \nu}+ \mathcal{T}_{\mu \nu})\,,
\end{equation}
where the complicated nonlocal aspects of the gravitational interaction are contained in the non-symmetric tensor $\mathcal{T}_{\mu \nu}$~\cite{BiMa}. The detailed form of $\mathcal{T}_{\mu \nu}\,K^\mu K^\nu$ is unknown at the present time; therefore, the implications of the Raychaudhuri 
equation for the occurrence of spacetime singularities in nonlocal gravity cannot be ascertained. On the other hand, it is possible to discuss the occurrence of cosmological singularities within the context of Newtonian cosmology~\cite{CT, KY}. The purpose of this section is to illustrate this result and discuss its straightforward extension to nonlocal Newtonian cosmology. 

To discuss the mathematical implications of our model in a general way, it is useful to write the main equations of our nonlocal Euler--Poisson dust model as 
\begin{align}\label{epg}
\nonumber \rho_t+\nabla \cdot (\rho u)&=0,\\
\nonumber u_t+(u\cdot \nabla) u&= - k \nabla \Phi,\\
 \Delta \Phi&=f(\rho),
\end{align}
where the dynamical variables, which are functions of time and space,   are usually taken to be the density  $\rho$ and velocity $u$ of some perfect fluid substance. The auxiliary quantity  $\Phi$  is a potential given by Poisson's equation where the source is some  \emph{positive} function $f$ of the density, which in the case under consideration would be
\begin{equation}\label{R3}
 f(\rho) = 4\pi G\,(\rho + \rho_D)\,,
\end{equation}
and the constant $k$ is chosen according to the application. The sign of $k$ determines the nature of the Newtonian force: In case $k$ is positive, the force is attractive;  when $k<0$, it is repulsive. In the case under consideration in this paper $k=1$, but for the sake of generality we leave $k \ne 0$ arbitrary in this section; similarly,  we will suppose that our dynamical variables are defined on 
$n$-dimensional Euclidean space. 

In the notation that we employ here, the material derivative of the velocity vector,
$\partial_t \,\mathbf{v} +(\mathbf{v}\cdot \nabla)\, \mathbf{v}$,
may be written in the form
\begin{equation}\label{R4} 
 u_t +(u\cdot \nabla) u\,;
\end{equation}
moreover, $\check{V}:=\nabla u$ is a matrix with components $\check{V}_{ij}=\partial_iv_j$. Under the assumption that the dynamical variables and the gravitational potential are at least twice continuously differentiable, a fruitful idea is to differentiate the momentum balance equation with respect to the spatial variables. The spatial derivative of the momentum balance equation is 
\begin{equation}\label{R5}
\frac{\partial}{\partial t} \check{V}+(u\cdot \nabla)\check{V}+(\check{V})^2 =-k \Hess \Phi\,,
\end{equation}
where $\Hess \Phi$ is the tidal matrix. The transpose of this equation is 
\begin{equation}\label{R6}
\frac{\partial}{\partial t} \check{V}^T+(u\cdot \nabla)\check{V}^T+(\check{V}^T)^2 =-k \Hess \Phi\,.
\end{equation}
Let us define $\mathcal{L}$ to be the (infinitesimal) strain rate matrix and $\Omega$ to be the vorticity matrix; then, 
\begin{equation}\label{cpn}
 \mathcal{L}:=\frac{1}{2} (\check{V}+\check{V}^T), \qquad \Omega:=\frac{1}{2} (\check{V}-\check{V}^T)\,
\end{equation}
and Eqs.~\eqref{R5} and~\eqref{R6} are equivalent to a system of two equations involving 
$\mathcal{L}$ and $\Omega$, namely, 
\begin{align}\label{bas}
\nonumber \mathcal{L}_t+(u\cdot \nabla)\mathcal{L}+\frac{1}{2} (\check{V}^2+(\check{V}^T)^2)&=-k \Hess \Phi\,,\\
 \Omega_t+(u\cdot \nabla)\Omega+\frac{1}{2} (\check{V}^2-(\check{V}^T)^2)&=0\,.
\end{align}
It follows from a simple calculation that
\begin{align}\label{scs}
\frac{1}{2} (\check{V}^2+(\check{V}^T)^2)&=\mathcal{L}^2 + \Omega^2\,,\\
\frac{1}{2} (\check{V}^2-(\check{V}^T)^2)&=\mathcal{L} \Omega+\Omega \mathcal{L}\,.
\end{align}
Thus, the desired equivalent system is 
\begin{align}\label{desn}
\nonumber  \mathcal{L}_t+(u\cdot \nabla)\mathcal{L}+\mathcal{L}^2 + \Omega^2 &=-k \Hess \Phi\,,\\
\Omega_t+(u\cdot \nabla)\Omega+\mathcal{L} \Omega+\Omega \mathcal{L}&=0\,.
\end{align}
The second equation in display~\eqref{desn} is linear in the vorticity matrix $\Omega$. Under the assumption that solutions are unique, this leads to a standard observation: If $\Omega$ vanishes for some $t$, then $\Omega$ vanishes for all $t$. Henceforward, we assume that $\Omega=0$.

Define $w$ to be the volume expansion such that  $w=\trace \mathcal{L}$ and note that $w$ is the divergence of the velocity field $u$. Using this notation and with the $n\times n$ identity matrix denoted by $I$, 
define the volumetric strain rate matrix $\mathcal{V}:= w I/n$ and the  shear strain rate matrix
\begin{equation}\label{R7} 
 \mathcal{S}:=\mathcal{L}-\mathcal{V}\,. 
\end{equation}
Using the new quantities, the momentum balance equation in the absence of vorticity takes the form 
\begin{equation}\label{sveq}
\mathcal{S}_t+(u\cdot \nabla)\mathcal{S}+\mathcal{V}_t+(u\cdot \nabla)\mathcal{V}+\mathcal{S}^2+\mathcal{S}\mathcal{V}+\mathcal{V}\mathcal{S}+\mathcal{V}^2=-k \Hess \Phi\,.
\end{equation}
In the decomposition $\mathcal{L}= \mathcal{S}+\mathcal{V}$,   $\mathcal{V}$ is a diagonal (hence symmetric) matrix such that
\begin{equation}\label{R8}  
 \trace{\mathcal{V}}=\trace{\mathcal{L}}, \qquad \trace{\mathcal{V}^2}=\frac{1}{n} w^2\,
\end{equation}
and  $\mathcal{S}$ is a symmetric matrix with zero trace. In general, the trace of the square of a traceless symmetric matrix is nonnegative. 

Suppose that $\Omega=0$ and consider the trace of Eq.~\eqref{sveq} along a Lagrangian trajectory $t\mapsto X(t,\xi)$, where $X$ is the Lagrangian flow map; that is, $X_t(t,\xi)=u(t, X(t,\xi))$. Here, $\xi$ could be the Lagrangian marker of a dust particle at some initial time $t=t_{in}$; that is, $X(t_{in},\xi)=\xi$.  
Taking the trace of both sides of Eq.~\eqref{sveq} and using Eq.~\eqref{Z4}, we find the family of ODEs (parametrized by the Lagrangian marker) given by
\begin{equation}\label{aaa}
\frac{\partial}{\partial t} w(t, X(t,\xi))+\frac{1}{n} w^2(t, X(t,\xi))  =-k f(\rho(t, X(t,\xi)))-\trace (\mathcal{S}^2(t, X(t,\xi)))\,.
\end{equation}
Here, we have used the linearity of the trace operator and  the properties of $\mathcal{S}$ and $\mathcal{V}$. Equation~\eqref{aaa}  could be viewed as the Euler--Poisson analog of the Raychaudhuri equation for congruences of geodesics on a Lorentzian manifold,  which is used extensively in the theory of singularity formation in cosmology~\cite{HE}. 

It proves useful to define
\begin{equation}\label{R9}  
W(t, \xi) :=w(t, X(t,\xi))\,, \qquad  \Sigma(t, \xi) := k f(\rho(t, X(t,\xi)))+\trace(\mathcal{S}^2(t, X(t,\xi)))\,,
\end{equation}
where $\rho(t, X(t,\xi))=\varrho (t, \xi)$ and 
\begin{equation}\label{R10}
W(t, \xi) = -\frac{1}{\varrho}\,\frac{\partial \varrho}{\partial t}\,.
\end{equation}
Equation~\eqref{aaa} may be written in the more compact form as
\begin{equation}\label{R11}
\frac{\partial W}{\partial t} = -\frac{1}{n}\, W^2 -\Sigma(t, \xi)\,,
\end{equation}
where $\Sigma(t, \xi) \ge0$. The ODE~\eqref{R11} is a scalar Riccati equation that exhibits shell-crossing singularities. 
If at some initial time $t_{in}$, $W(t_{in}, \xi)>0$, then Eq.~\eqref{R11} implies that the solution of this equation in reverse time is eventually singular; that is, at some time $t$, $t<t_{in}$, $W=+\infty$.  Similarly, if at some initial time   
$t_{in}$, $W(t_{in}, \xi)<0$, then it follows from Eq.~\eqref{R11} that eventually in the future $W \to -\infty$.   These singularities, in the context of the standard cosmological models,  correspond to the Big Bang and Big Crunch singularities, respectively. The position-dependent shell-crossing singularities  of Eq.~\eqref{R11} occur in finite time in the past or the future. This can be simply seen from the fact that 
\begin{equation}\label{R12}
\frac{\partial \mathcal{W}}{\partial t} = -\frac{1}{n}\, \mathcal{W}^2\,
\end{equation}
has the solution 
\begin{equation}\label{R13}
\mathcal{W}(t, \xi)  = \frac{n}{t-t_{in} +\frac{n}{\mathcal{W}(t_{in}, \,\xi)}}\,. 
\end{equation}
It is intuitively clear that with $\Sigma(t, \xi) \ge0$, the finite intervals of time to the shell-crossing  singularities, which can be estimated using Eq.~\eqref{R11}, are actually shorter than those implied by Eq.~\eqref{R13}. A sharper result in connection with the occurrence of singularities in finite time is contained in Ref.~\cite{KY}.

\section{DISTENTION}

The essential difference between nonlocal gravity and general relativity is that in nonlocal gravity the gravitational field is history dependent. This difference can be equivalently expressed in terms of spacetime memory or the presence of effective dark matter. 

Memory fades in space and time and this leads to a crucial assumption in nonlocal Newtonian cosmology, namely, the fraction of the effective dark matter to baryonic matter $f_{DM}$ decreases with the expansion of the universe, since memory fades with cosmic time. Indeed, if $A(t)$ is the expansion scale factor, then we have assumed that $f_{DM} \propto A^{-\varpi}$ with $\varpi >0$. This means that 
the attraction of gravity gradually decreases toward the standard Newtonian form, since $f_{DM}$ can eventually decrease to zero.  As the amount of effective dark matter decreases with cosmic time, the strength of the gravitational attraction also decreases and the process of  structure formation in the universe slows down. Another consequence of the decline in the attractive force of gravity involves the evolution of structures that have already formed. Indeed,  
an isolated self-gravitating system could experience distention as the universe expands, since the strength of the internal gravitational attraction of the system gradually decreases with cosmic time as well.  Such a dilation would be in addition to the tidal influence of the rest of the expanding universe on the isolated system~\cite{Mashhoon:2007qm}. It is possible that such a dilation mechanism is in part responsible for the significant size evolution of high-mass quiescent early-type galaxies that show no evidence of recent star-formation activity. The dilation of such early-type galaxies has been observationally established from $A \approx 0.3$ to $A \approx 1$---see, for instance~\cite{VDOK1, VDOK2, DAM}  and the references cited therein. 

Let us assume that in Eqs.~\eqref{CM1} and~\eqref{CM2}, $\rho (t, \mathbf{x}) = M \delta(\mathbf{x})$, which represents a Dirac delta function source of mass $M$ at the origin of the Cartesian coordinate system; then, using Eqs.~\eqref{CM1}--\eqref{CM3} and~\eqref{N7}, it is possible to derive the ``Newtonian" gravitational force on a point mass $m$ due to the point mass $M$ taking due account of cosmic evolution. The result is
\begin{equation}\label{DI1}
\mathcal{F}_{NLG}(t, \mathbf{r}) = -GmM\,\frac{\widehat{\mathbf{r}}}{r^2}\,\left\{ 1+A(t)^{-\varpi}\,[\alpha_0- \alpha_0\,(1+\frac{1}{2}\,\mu_0\,r)\,e^{-\mu_0\,r}-\mathcal{E}(r)] \right\}\,,
\end{equation}
where $\mathbf{r}$ is the vector that extends from $M$ to $m$, $\widehat{\mathbf{r}}= \mathbf{r}/r$ and $\mathcal{E}$ is given by
\begin{equation}\label{DI2}
 \mathcal{E}_1(r)=\frac{a_0}{\lambda_0}e^{\varsigma}\Big[E_1(\varsigma)-E_1(\varsigma+\mu_0 r)\Big]\,
\end{equation}
or
\begin{equation}\label{DI3}
 \mathcal{E}_2(r)= 2\, \mathcal{E}_1(r) - \frac{a_0}{\lambda_0}\,\frac{r}{r+a_0}e^{-\mu_0 r}\,,
\end{equation}
depending on whether we employ reciprocal kernel $q_1$ or $q_2$ in Eq.~\eqref{CM2}, cf.~\cite{Mashhoon:2014jna}. Here, we have introduced new parameters $\alpha_0$ and $\varsigma$,
\begin{equation}\label{DI4}
\alpha_0 = \frac{2}{\lambda_0\,\mu_0}\,, \qquad \varsigma= \mu_0\,a_0\,,
\end{equation}
and the \emph{exponential integral function}~\cite{A+S}
\begin{equation}\label{DI5}
E_1(x):=\int_{x}^{\infty}\frac{e^{-t}}{t}dt\,.
\end{equation}
For $x: 0 \to \infty$,  $E_1(x)$ is a  positive function that monotonically decreases from infinity to zero. Moreover,  $E_1(x)$  behaves like $-\ln x$ near $x=0$ and vanishes exponentially as $x \to \infty$. In fact,
\begin{equation}\label{DI6}
E_1(x)=-C_E-\ln x -\sum_{n=1}^{\infty}\frac{(-x)^n}{n~ n!}\,,
\end{equation}
where $C_E=0.577\dots$ is Euler's constant. It can be shown that 
$\mathcal{E}_1(r)$ and $\mathcal{E}_2(r)$ are monotonically increasing positive functions of $r$ that start from zero at $r=0$ and for $r \to \infty$ asymptotically approach 
\begin{equation}\label{DI7}
\mathcal{E}_2(\infty)= 2\,\mathcal{E}_1(\infty)=\alpha_0\, \varsigma\,e^{\varsigma}E_1(\varsigma)<\alpha_0\,.
\end{equation}
It follows that as $r \to \infty$, the two-body force~\eqref{DI1} approaches
\begin{equation}\label{DI8}
\mathcal{F}_{NLG}(t, \mathbf{r}) \approx -\frac{GmM\,[1+ \tilde{\alpha}_0\,A^{-\varpi}(t)]}{r^2}\,\widehat{\mathbf{r}}\,,
\end{equation} 
which is a ``Newtonian" inverse square law with $G \to \tilde{G}(t) = G\,[1+ \tilde{\alpha}_0\,A^{-\varpi}(t)]$. Here,
\begin{equation}\label{DI9}
\tilde{\alpha}_0:=\alpha_0\,  \epsilon(\varsigma) >0\,,
\end{equation}
where
\begin{equation}\label{DI10}
\epsilon_1(\varsigma) = 1-\frac{1}{2}\,\varsigma\,e^\varsigma\,E_1(\varsigma)\,, \qquad \epsilon_2(\varsigma) = 1-\varsigma\,e^\varsigma\,E_1(\varsigma)\,
\end{equation}
depending on whether we employ reciprocal kernel $q_1$ or $q_2$. Another useful way to interpret  Eq.~\eqref{DI8} is that if $a_0=0$, the net effective dark mass associated with point mass $M$ at the present epoch ($A=1$) is simply $\alpha_0\,M$, where $\alpha_0 \approx 11$; on the other hand, for $a_0 \ne 0$, the corresponding result is somewhat smaller and given by $\alpha_0\, \epsilon_i(\varsigma)\, M$, for $i=1,2$.

The Newtonian inverse square force law satisfies the shell theorem, namely, a homogeneous spherical distribution of matter attracts an external particle as if the mass of the sphere were concentrated at its center; moreover, in the hollow interior of a homogeneous spherical shell of matter, there is no force of gravity. These results are related in general relativity to Birkhoff's theorem, which is not expected to hold in nonlocal gravity. Indeed, the force law~\eqref{DI1} violates the shell theorem as described in detail in Appendix A.  

Equation~\eqref{DI1} consists of two distinct parts: a standard Newtonian inverse square part and an effective dark matter part whose strength monotonically decreases with the expansion of the universe. Isolated $N$-body self-gravitating systems are held together by attractive gravitational forces of the form given by Eq.~\eqref{DI1}; however, the manner in which distention occurs  in an $N$-body system such as a galaxy is a difficult problem that is beyond the scope of this work. To get some idea of this dynamical evolution, we consider a bounded two-body system in the rest of this section. 

The analog of the inverse square law of force in the Newtonian limit of nonlocal gravity, namely, 
Eq.~\eqref{DI1} at the present epoch ($A=1$), essentially behaves as $1/r^2$ for $r \to 0$ and $r \to \infty$; furthermore, in the intermediate Tohline--Kuhn regime~\cite{Mashhoon:2007qm}, namely, for $a_0<r<\mu_0^{-1}$, it behaves as $1/r$ in agreement with the ``flat" rotation curves of spiral galaxies~\cite{RaMa}. At any given cosmic epoch $t$, the standard two-body problem with central force~\eqref{DI1} is such that the relative orbit is planar; moreover, the corresponding effective potential has qualitatively the same form as the effective potential in the Kepler system. Thus there are stable circular orbits, bound orbits with two apsidal distances corresponding to the turning points of the effective potential as well as scattering orbits.  

To simplify matters even further, we will consider the distention problem for an attractive central force of the form
\begin{equation}\label{DI11}
\mathcal{F}(t, \mathbf{r}) = -\frac{\gamma(t)}{r^n}\,\widehat{\mathbf{r}}\,,
\end{equation} 
where $\gamma(t) >0$ and $n<3$. As is well known, for $n<3$ this central force admits stable circular orbits.  Beyond these circular orbits, there are bound orbits with apsidal distances $\mathbb{A}_n$ and $\mathbb{B}_n$. The variation of $\gamma$ with cosmic time is given by
\begin{equation}\label{DI12}
\gamma(t) =\frac{\gamma_0}{A^{\varpi}(t)}\,, 
\end{equation} 
where $\gamma_0:=\gamma(t_0)>0$ is a constant. 

Consider a stable circular orbit of radius $r_c$ given by
\begin{equation}\label{DI13}
r_c= \left(\frac{\gamma}{\mathbb{L}^2}\right)^{\frac{1}{n-3}}\,, 
\end{equation} 
where $\mathbb{L}$ is the orbital angular momentum of the orbit per unit mass. We note that the temporal variation of $\gamma$ is very slow in comparison with the fast circular motion; therefore, as $\gamma$ changes slowly with cosmic time, $\mathbb{L}$ remains the same and the size of the circular orbit varies as 
\begin{equation}\label{DI14}
r_c \propto \gamma^{\frac{1}{n-3}}\,. 
\end{equation} 
We conjecture that this relation is valid not just for the radius of a circular orbit but for the size of any bound orbit within the framework of the two-body system under consideration here.  That is,
\begin{equation}\label{DI15}
{\rm orbital~dimensions}  \propto \gamma^{\frac{1}{n-3}}\,. 
\end{equation}  
Preliminary numerical investigations point to the validity of this conjecture. We next consider the proof of this conjecture for  two simple cases involving $n=2$ and $n=-1$, respectively. 

For the case of $n=2$, namely, the Newtonian inverse square law of attraction with time varying gravitational force, our conjecture implies that the size of the orbit would increase in proportion to $1/\gamma$---cf. Eq.~\eqref{DI15} for $n=2$ . The adiabatic invariance of the action variables has been discussed in Ref.~\cite{LL}. It follows from the adiabatic invariants of the Keplerian two-body system that if the strength of the attractive force decreases very slowly, the orbit keeps its shape, but its size increases in accordance with our conjecture. That is, the eccentricity of the orbit remains the same, while its dimensions increase in inverse proportion to the strength of the force~\cite{LL}. The Kepler system with slowly decreasing mass is further discussed in Appendix B, where the cosmological evolution of the system is treated in detail and the importance of averaging is emphasized.

Let us next consider $n=-1$; in this case, we have a linear restoring force and our conjecture  implies that the size of the orbit would increase in proportion to $\gamma^{-\frac{1}{4}}$---cf. Eq.~\eqref{DI15} for $n=-1$. Let us assume for the moment that $\gamma$ is constant. The planar orbit in this well-known case corresponds to simple harmonic oscillations along the Cartesian $x$ and $y$ axes of the orbital plane with the same frequency $\sqrt{\gamma}$. The orbit is an ellipse in this case as well with semimajor axis $\mathbb{A}_{-1}$ and semiminor axis $\mathbb{B}_{-1}$; moreover,  the corresponding action variables can be simply computed. It follows from the adiabatic invariance of the action variables that as $\gamma$ decreases very slowly with time, 
$\mathbb{A}_{-1}\, \gamma^{\frac{1}{4}}$ and $\mathbb{B}_{-1}\, \gamma^{\frac{1}{4}}$
remain unchanged, in agreement with our conjecture. 

\section{Structure Formation: Spherical Symmetry}

We are interested here in the formation of spherically symmetric structure on the homogeneous and isotropic background described in Section II, which is consistent with nonlocal Newtonian cosmology under consideration in this paper. In this section, we return to the Eulerian formulation of our dust model contained in Eqs.~\eqref{Z1} and~\eqref{Z2}.  

We assume that the center of the perturbation coincides with the origin of the background Cartesian coordinates $\mathbf{x}$. Thus the dust density is $\rho=\rho(t, r)$ and its radial velocity is given by 
$v=v(t, r)$, where $r=|\mathbf{x}|$ is the radial coordinate. Using subscript notation for partial derivatives, it is straightforward to show that 
the Euler-Poisson system under the assumption of spherical symmetry reduces to
\begin{equation}\label{F1}
\rho_t+  (\rho\, v)_r+ \frac{2}{r}\, \rho\, v=0\,
\end{equation}
and 
\begin{equation}\label{F2}
(v_t+v_r\,v)_r +  \frac{2}{r}\, (v_t+v_r\,v)= -4 \pi G\, (\rho + \rho_D)\,,
\end{equation}
where $\rho_D$ is given by Eq.~\eqref{CM2}.

Let us first check that the spatially homogeneous and isotropic cosmological background~\eqref{N5} is a solution of Eqs.~\eqref{F1} and~\eqref{F2}.  To this end, let $\rho = \bar{\rho}$ and $v=\bar{v}$, where
\begin{equation}\label{F3}
\bar{\rho} = \frac{\rho_0}{A^3}\,, \qquad \bar{v} = H(t)\, r\,,
\end{equation}
as in Eq.~\eqref{N5}. The Hubble parameter is defined by
\begin{equation}\label{F4}
H= \frac{\dot A(t)}{A}\,,
\end{equation}
where, 
\begin{equation}\label{F5}
\ddot A =-\frac{4\pi G \rho_0}{3\,A^{2}} \,[1+\frac{\tilde{\alpha}_0}{A^\varpi}]\,
\end{equation}
and
\begin{equation}\label{F6}
H^2= \frac{8\pi G\rho_0}{3} \left[\frac{1}{A^3}+\frac{\tilde{\alpha}_0}{(\varpi +1)\, A^{\varpi +3}}\right]\,,
\end{equation}
in agreement with Eq.~\eqref{N8} with $\bar{E}=0$. Moreover, we note that 
\begin{equation}\label{F7}
\dot H = \frac{\ddot A}{A}-H^2 = -\frac{4\pi G \rho_0}{A^{3}} \,[1+\frac{\varpi +3}{3\,(\varpi + 1)}\,\frac{\tilde{\alpha}_0}{A^\varpi}]\,.
\end{equation} 
It is simple to verify that Eqs.~\eqref{F1} and~\eqref{F2} are indeed satisfied, as expected. 

In practice, we are interested in the evolution of density inhomogeneities during the expansion of the universe. Therefore, in Eqs.~\eqref{F1} and~\eqref{F2}  we let
\begin{equation}\label{F8}
\rho = \bar{\rho}\,(1 + \mathbb{D})\,, \qquad v = \bar{v}\,(1 +\mathbb{V})\,.
\end{equation}
Then, we must solve the nonlocal and nonlinear perturbation equations for the \emph{density contrast} $\mathbb{D}$ and \emph{radial velocity contrast} $\mathbb{V}$.  This will be done  \emph{numerically}  in the next section. The rest of the present section is devoted to the solution of the \emph{linearized} perturbation equations for $|\mathbb{D}| \ll 1$ and $|\mathbb{V}| \ll 1$ in order to gain insight into the nature of the numerical results presented in Section VII.

\subsection{Local Model with $\rho_D=0$}

The \emph{local} model is obtained by setting $\alpha_0 = \tilde{\alpha}_0=0$. In the \emph{absence of nonlocality}, 
Eqs.~\eqref{F1} and~\eqref{F2} reduce to the form
\begin{equation}\label{F9}
\check{\rho}_t+  (\check{\rho}\, v)_r=0\,
\end{equation}
and 
\begin{equation}\label{F10}
[r^2\,(v_t+v_r\,v)]_r = -4 \pi G\, \check{\rho}\,,
\end{equation}
where $\check{\rho}=r^2 \rho$. The homogeneous and isotropic solution of this local model is given by Eq.~\eqref{F3}, where the corresponding background scale factor of the expanding universe is given by
\begin{equation}\label{F11}
A(t) = (t / t_0)^{2/3}\,, \qquad t_0 = (6 \pi G\, \rho_0)^{-1/2}\,.
\end{equation}

Let us first consider the linearization of Eqs.~\eqref{F9} and~\eqref{F10} in terms of $\mathbb{D}$ and $\mathbb{V}$ using Eq.~\eqref{F8}. Moreover, we change the temporal and radial variables to dimensionless variables $\bar{s}$ and $\bar{r}$, respectively, where
\begin{equation}\label{F12}
A(t) = \exp{{\bar{s}}}\,, \qquad r=\frac{1}{\mu_0}\,  \exp{{\bar{r}}}\,.
\end{equation}
The linearized form of Eqs.~\eqref{F9} and~\eqref{F10} can now be written for $|\mathbb{D}| \ll 1$ and $|\mathbb{V}| \ll 1$ as
\begin{equation}\label{F13}
\mathbb{D}_{\bar{s}} + \mathbb{D}_{\bar{r}}+ \mathbb{V}_{\bar{r}}+ 3\,\mathbb{V}=0\,
\end{equation}
and 
\begin{equation}\label{F14}
\mathbb{D} + \mathbb{V}+ 2\, \mathbb{V}_{\bar{s}}+ \frac{7}{3}\, \mathbb{V}_{\bar{r}}+ \frac{2}{3}\,(\mathbb{V}_{\bar{s} \bar{r}} + \mathbb{V}_{\bar{r} \bar{r}})=0\,.
\end{equation} 
We look for solutions of the form 
\begin{equation}\label{F15}
\mathbb{D} = \mathbb{D}_0\,\exp (\gamma_1\,\bar{s}+\gamma_2\,\bar{r})\,, \qquad \mathbb{V}=\mathbb{V}_0\,\,\exp (\gamma_1\,\bar{s}+\gamma_2\,\bar{r})\,,
\end{equation}
where $\mathbb{D}_0$, $\mathbb{V}_0$, $\gamma_1$ and $\gamma_2$ are constants. This means that $\mathbb{D}$ and $\mathbb{V}$ are linear combinations of terms of the form $A^{\gamma_1}\,(\mu_0 \, r)^{\gamma_2}$ with constant coefficients.  The density contrast and its radial derivative are expected to be finite at the origin $r=0$; in fact, this requirement means that either
\begin{equation}\label{F15a}
 \gamma_2 = 0\,, \qquad  {\rm or}~~~ \gamma_2 \ge 1\,.
\end{equation}
It follows from Eqs.~\eqref{F13} and~\eqref{F14} that 
\begin{equation}\label{F16}
(\gamma_1 + \gamma_2)\,\mathbb{D}_0 + (\gamma_2 + 3)\,\mathbb{V}_0=0\,
\end{equation}
and 
\begin{equation}\label{F17}
\mathbb{D}_0 + \left(1+2\,\gamma_1 + \frac{7}{3}\,\gamma_2 + \frac{2}{3}\,\gamma_1\,\gamma_2+\frac{2}{3}\,\gamma_{2}^2\right)\,\mathbb{V}_0=0\,.
\end{equation}
These equations have solutions provided the corresponding determinant vanishes; that is, 
\begin{equation}\label{F18}
(\gamma_1 + \gamma_2)\, \left(1+2\,\gamma_1 + \frac{7}{3}\,\gamma_2 + \frac{2}{3}\,\gamma_1\,\gamma_2+\frac{2}{3}\,\gamma_{2}^2\right)= \gamma_2 +3\,.
\end{equation}
It follows from Eq.~\eqref{F15a} that $\gamma_2 \ne -3$; therefore, relation~\eqref{F18} simplifies to
\begin{equation}\label{F18a}
2\,(\gamma_1 + \gamma_2)^2 + (\gamma_1 + \gamma_2)-3 =0\,.
\end{equation}
Hence, 
\begin{equation}\label{F18b}
\gamma_1 + \gamma_2 = 1\,, \qquad {\rm or}~~ ~\gamma_1 + \gamma_2 = -\frac{3}{2}\,.
\end{equation}
Putting Eqs.~\eqref{F15a} and~\eqref{F18b} together, we find that when $\gamma_2=0$ we have either $\gamma_1=1$ or  $\gamma_1=-3/2$.
 If follows that in this case the density contrast  is proportional to a linear combination of $A$ and  $A^{-3/2}$, which are the usual growing and decaying modes of the standard linearized theory~\cite{Chicone:2015sda, Pe, ZN, Mu}. Alternatively, $\gamma_2 \ge 1$ and then Eq.~\eqref{F18b} implies that  $\gamma_1 \le 0$, which rules out a growing mode. \emph{Thus the main result of the local linearized theory is that there is a unique growing mode that is independent of the radial distance; this mode will be dominant over time and corresponds to $\gamma_1=1$ and $\gamma_2=0$.}

Concentrating on the growing mode, we note that with $\gamma_1=1$ and $\gamma_2=0$, the linear perturbation equations imply that $\mathbb{D}$ and $\mathbb{V}$  are simply given by $\mathbb{D} = -3\,\mathbb{V}= -3\, \mathbb{V}_0\,A$. It follows from this solution for the growing mode that if we assume, for instance,  that at the epoch of decoupling $A(t_{dec}) = 10^{-3}$ and $\mathbb{D}(t_{dec}) = 10^{-5}$, then at the present epoch we find
$\mathbb{D}(t_0) = 10^{-2}$, which is not sufficient for structure formation. 

Next, we turn to the \emph{local nonlinear} Eqs.~\eqref{F9} and~\eqref{F10}. Except for the background solution with $\mathbb{D} = \mathbb{V}=0$, it does not seem possible to solve these equations analytically for $\mathbb{D}$ and $\mathbb{V}$; therefore, we resort to the numerical integration of  Eqs.~\eqref{F9} and~\eqref{F10}. For initial conditions, we assume as before that at the era of decoupling  $A(t_{dec})=10^{-3}$ and the density contrast  $\mathbb{D}$ is a Gaussian with amplitude $\delta = 10^{-5}$; that is,   
\begin{equation}\label{F19}
\mathbb{D}(t_{dec}, r)= \delta \,  \exp\left({-\frac{\hat{r}^2}{2\,\hat{w}^2}}\right)\,, \qquad  \mathbb{V}(t_{dec}, r)=0\,,
\end{equation}
where $\hat{r} = \mu_0\,r$ and  $\hat{w}$ is a dimensionless measure of the width of the Gaussian perturbation. In fact, we have
\begin{equation}\label{F20}
\sqrt{\frac{2}{\pi}}\, \int_0^{\infty}e^{-\frac{\hat{r}^2}{2\,\hat{w}^2}}\, d\hat{r} = \hat{w}\,. 
\end{equation}
Figure 3 illustrates the result of numerical integration of Eqs.~\eqref{F9} and~\eqref{F10}. It is interesting to compare the results of the nonlinear theory with those of the linearized theory, where for the growing mode we have at the present epoch $A(t_0)=1$,
 $\mathbb{D}=-3\,\mathbb{V}$ and $\mathbb{D} =0.01$. These theoretical results for the linearized local theory are in reasonable agreement with the results given in Figure 3 that represent the numerical integration of the local nonlinear equations of our spherically symmetric model. This concordance is  due to the fact that the Gaussian perturbation has an amplitude that is always rather small compared to unity.   
 
\begin{figure}
\begin{center}
\includegraphics[width=7.5 cm]{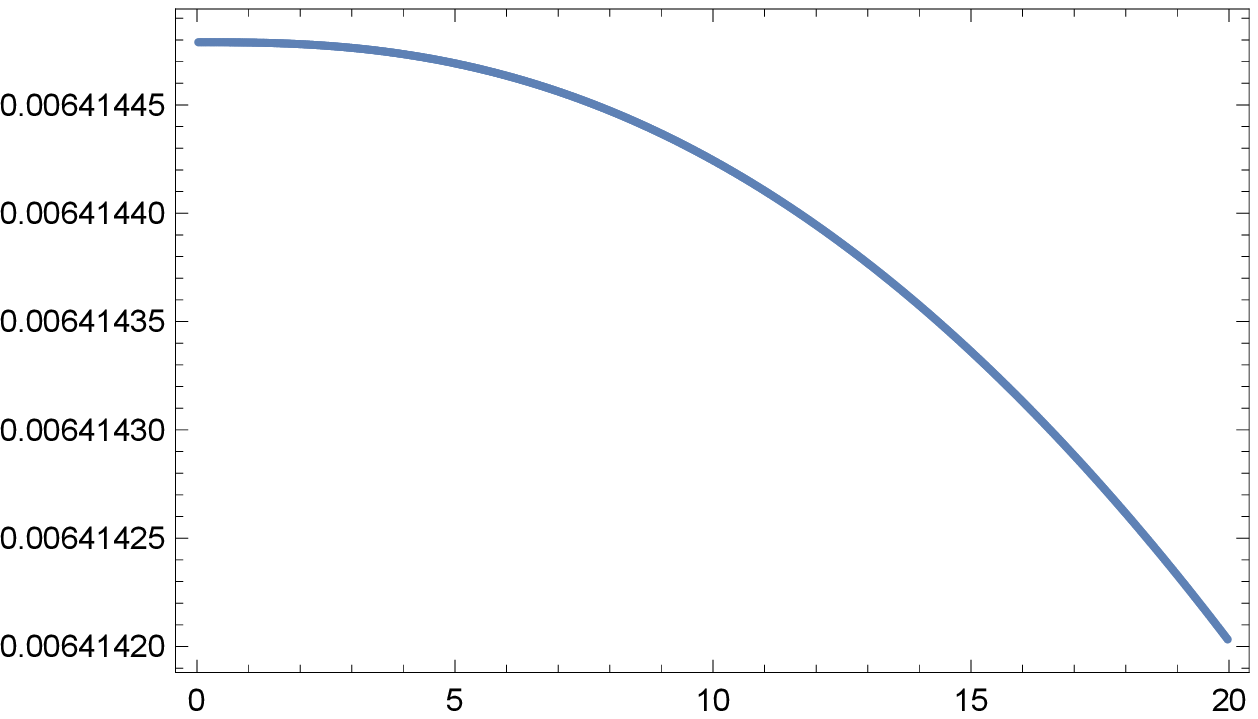} \quad  \includegraphics[width=7.5 cm]{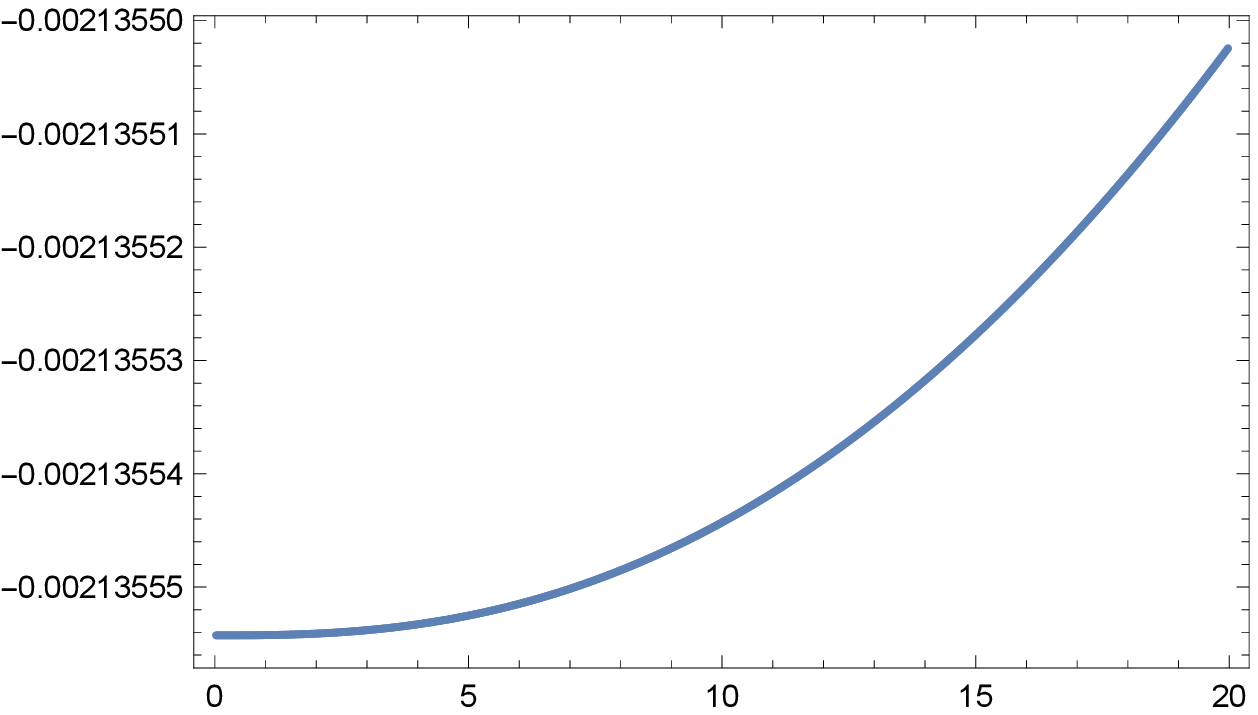}
\end{center}
\caption{
The left panel is a numerically generated graph of the density contrast $\mathbb{D}$ versus radial distance $\hat{r}$ for the local dust model at the moment that the scale factor $A$ reaches unity after starting with $A(t_{in})=10^{-3}$. The initial density contrast is given by the Gaussian $\delta\, \exp{[-\hat{r}^2/(2 \hat{w}^2)]}$ with $\delta=10^{-5}$, $\hat{r} = \mu_0\,r$ and $\hat{w}=2$.  The initial radial velocity contrast is set to zero. The right panel depicts the radial velocity contrast $\mathbb{V}$ for the same numerical experiment.
\label{fig3} }
\end{figure}
 
\subsection{$\rho_D=(\tilde{\alpha}_0 / A^\varpi)\, \rho$}

Let us return to the Euler--Poisson system given by Eqs.~\eqref{F1} and~\eqref{F2}. Even in linearized perturbation theory for $\mathbb{D}$ and $\mathbb{V}$, the analytic solution of this system  does not appear tractable. Therefore, we consider a useful limiting situation involving an initial perturbation in baryonic density that is nearly constant over the largest possible scale in radial distance. On such large spatial scales that would persist over cosmic epochs, we expect that $\rho_D$ is essentially equal to $(\tilde{\alpha}_0 / A^\varpi)\, \rho$, where the exact relation is valid for the uniform density case as in Eq.~\eqref{N4}. Therefore, we replace Eq.~\eqref{F2} with the relation
\begin{equation}\label{F21}
(v_t+v_r\,v)_r +  \frac{2}{r}\, (v_t+v_r\,v)= -4 \pi G\, \left(1+ \frac{\tilde{\alpha}_0}{A^\varpi}\right)\,\rho\,.
\end{equation}
Next, treating $\mathbb{D}$ and $\mathbb{V}$ to linear order, we obtain as before Eq.~\eqref{F13}, which is the linearized continuity equation in this case, and 
\begin{equation}\label{F22}
\mathbb{I}\, \mathbb{D} + (2-\mathbb{I})\, \mathbb{V}+ 2\, \mathbb{V}_{\bar{s}}+ \frac{1}{3}\,(8-\mathbb{I})\, \mathbb{V}_{\bar{r}}+ \frac{2}{3}\,(\mathbb{V}_{\bar{s} \bar{r}} + \mathbb{V}_{\bar{r} \bar{r}})=0\,,
\end{equation} 
where 
\begin{equation}\label{F23}
\mathbb{I} := \frac{1+  \tilde{\alpha}_0\,\exp(-\bar{s}\,\varpi)}{1+  \frac{\tilde{\alpha}_0}{\varpi +1}\,\exp(-\bar{s}\,\varpi)}\,. 
\end{equation}
For $ \tilde{\alpha}_0=0$, $\mathbb{I}=1$ and Eq.~\eqref{F22} reduces to Eq.~\eqref{F14}. 

In conformity with our basic approach here, we assume henceforth that $\mathbb{D}$ and $\mathbb{V}$ are essentially independent of radial distance; that is, we set
\begin{equation}\label{F24}
\mathbb{D}_{\bar{r}}= \mathbb{V}_{\bar{r}}=0\,.
\end{equation}
Then, Eqs.~\eqref{F13} and~\eqref{F22} reduce to 
\begin{equation}\label{F25}
\mathbb{V}=-\frac{1}{3}\, \mathbb{D}_{\bar{s}}\,
\end{equation}
and 
\begin{equation}\label{F26}
2\,\mathbb{D}_{\bar{s} \bar{s}} +(2-\mathbb{I})\,\mathbb{D}_{\bar{s}}-3\,\mathbb{I}\,\mathbb{D}=0\,,
\end{equation}
respectively. It is straightforward to show that Eq.~\eqref{F26} for $\mathbb{D}$ is exactly the same differential Eq.~\eqref{Z25} that $\Psi$ satisfies. This is a remarkable result: While in the Zel'dovich approach $\Psi$ is the Lagrangian density contrast both in the linear as well as the nonlinear regimes, $|\mathbb{D}| \ll 1$ is the Eulerian density contrast in our linear perturbation approach here. Thus, Eq.~\eqref{F26} implies that $\mathbb{D}$ is given by a linear combination of the growing mode $\mathcal{S}_{\sigma_{+}}(A)$ and the decaying mode $\mathcal{S}_{\sigma_{-}}(A)$ given by Eq.~\eqref{Z32}.

Adopting the growing mode for the growth of the density contrast with cosmic time, we find that for the problem of structure formation under consideration here,
\begin{equation}\label{F27}
\mathbb{D}(A)= 10^{-5}\,\frac{\mathcal{S}_{\sigma_{+}}(A)}{\mathcal{S}_{\sigma_{+}}(10^{-3})}\,
\end{equation} 
and 
\begin{equation}\label{F28}
\mathbb{V}(A)= -\frac{1}{3}\, A\, \frac{d\mathbb{D}}{dA}\,.
\end{equation} 
On large scales, e.g.\ for an initial Gaussian with a wide width, we expect that the numerical results of the exact nonlocal and nonlinear system would approach Eqs.~\eqref{F27} and~\eqref{F28} for 
$|\mathbb{D}| \ll 1$ and $|\mathbb{V}| \ll 1$.

\section{Numerical Experiments}

To approximate spherically symmetric solutions of the nonlocal dust model,  we employ Cartesian coordinates $(x,y,z)$  in space  and set  $r:=\sqrt{x^2+y^2+z^2}$. Under the assumption of spherical symmetry,  density  $\rho$  and velocity $\mathbf{v}$ are sought as  functions of  $t$ and $r$.   Velocity, as before,  is assumed to have the form $\mathbf{v}(t,r)=v(t,r)\, \widehat{ \mathbf{r}}$, where $\mathbf{r}$ is the usual position vector centered at the origin of the coordinate system and $\widehat{ \mathbf{r}}$ is the corresponding unit vector $\mathbf{r}/r$.  To avoid division by $r$ in some numerical computations,  the alternate formulation $\mathbf{v}(t,r)=V(t,r)\,  \mathbf{r}$  is used in this section.  Likewise, the \emph{gradient} of the gravitational potential $\Phi$ is the spherically symmetric vector field  $\Theta(t,r)\,\mathbf{r}$, where $\Phi_r=r\, \Theta$. Straightforward calculations show that,  for a general spherically symmetric vector field $w(t,r)= W(t,r)\, \mathbf{r}$ and using subscript notation for partial derivatives, the divergence operator produces
\[(\nabla \cdot w)(t,r)= r W_r(t,r)+3 W(t,r)\,.\]
Likewise, 
\[((w\cdot \nabla) w )(t,r)= (r W(t,r) W_r(t,r)+W(t,r)^2)\, \mathbf{r}\,.\]

Using these computations and the notation mentioned above, the dust model reduces to a system of three partial differential equations for the three unknown functions $\rho$, $V$, and $\Theta$:
\begin{align}\label{rdm}
\nonumber
\rho_t+ r (\rho V)_r+ 3 \rho V&=0\,,\\
\nonumber
V_t+r VV_r+V^2 &=-\Theta\,,\\
r\, \Theta_r+3 \Theta&=4 \pi G (\rho+\Gamma(\rho))\,,
\end{align}
where $\Gamma$ is the operator corresponding to the nonlocal characterization of the effective dark matter.  
In this formulation where the gradient of the gravitational potential is supposed to be computed directly,  the computed function $r\, \Theta$ must be proved to be  the gradient of a potential. In other words, the differential equation $\Phi_r=r\, \Theta$ must have a continuous solution defined up to an additive constant. For a continuous function $\mathbb{H}$,  the general solution of the ODE
\[ r \Theta_r+3 \Theta=\mathbb{H}(t,r)\,,\]
is given by
\[ 
r \Theta(t,r)=\frac{k_0}{r^2}+\frac{1}{r^2} \int_0^r \sigma^2 \mathbb{H}(t,\sigma)\,d\sigma,
\]
where $k_0$ is the constant of integration.  This constant must vanish for $r\, \Theta$ to remain bounded as $r\to 0$. In this case, the function $(t,r)\mapsto r\, \Theta(t,r)$ vanishes at $r=0$ and is continuous at this point for all $t$.   The desired gravitational potential is obtained by integration of $r\, \Theta(t,r)$.   Of course, the exact solution may be used directly to recast system~\eqref{rdm}  in the form 
\begin{align}\label{rdmrecast}
\nonumber
\rho_t+ r (\rho V)_r+ 3 \rho V&=0\,,\\
V_t+r VV_r+V^2 &=- 4 \pi G \frac{1}{r^3} \int_0^r \sigma^2 (\rho(t,\sigma)+\Gamma(\rho)(t,\sigma))\,d\sigma\,.
\end{align}

A formula for the nonlocal term is obtained by rewriting the integral in the definition of the nonlocal operator
\begin{equation}\label{orgam}\Gamma(g)(t,x)=\int q(t, \abs{x-y}) g(t,y)\,dy\end{equation}
in a suitable form for the spherical symmetry of the present model. The key ideas are to use spherical coordinates where the polar axis is in the direction of the vector pointing from the origin to $x$.  With this choice of coordinates, the quantity $\abs{x-y}$ depends only on the lengths of $x$ an $y$ and the angle $\theta$ between the corresponding position vectors. Because of the choice of polar axis, the integration over the corresponding azimuthal angle ($0 \to 2\,\pi$) can be carried out and the result is
\begin{align}\label{gfor}
\nonumber  \Gamma(\rho)(t,r)&=2\pi \int_0^\pi \int_0^\infty  q(t, \sqrt{r^2+\sigma^2-2 r\sigma \cos\theta}\,) \rho(t,\sigma) \sigma^2 \sin\theta\, d\theta d\sigma\\
&=2\pi \int_0^\infty \Big( \int_{-1}^1  q(t, \sqrt{r^2+\sigma^2-2 r\sigma X}\,) \,dX\Big) \rho(t,\sigma) \sigma^2\, d\sigma\,.
\end{align}

Let us first check that the spatially homogeneous and isotropic cosmological background~\eqref{N5} is a solution of system~\eqref{rdmrecast}. To this end, let 
\begin{equation}\label{B1}
\bar{\rho} = \frac{\rho_0}{A^3}\,, \qquad \bar{V} =\frac{\dot A}{A}\,.
\end{equation}
It is simple to verify that, as expected,  system~\eqref{rdmrecast} is satisfied provided Eq.~\eqref{F5} holds for $A(t)$.

For computation, an alternate scaling of the model equations is desirable. Recall that $\rho_0$ is the current baryonic density of the universe and $1/\mu_0$ is a galactic length of about 17 kpc. Also, with our choice of variables $V$ has dimensions of inverse time.  Using these parameters and with 
\[ \tilde t_0:= \sqrt{\frac{3}{4\pi G \rho_0}}\,, \]
we use the hatted dimensionless variables given by
\begin{equation}\label{newtdv}
 t=\tilde t_0\, \hat t\,,\quad  r=\frac{\hat r}{\mu_0}\,, \quad \rho=\rho_0\, \hat \rho\,, \quad V= \frac{\hat V}{\tilde t_0}\,.
\end{equation}
This choice of variables together with the change of variables $\hat \sigma=\mu_0\, \sigma$ in the integral transforms the model equations~\eqref{rdmrecast} to the form 
\begin{align}\label{sysdimless}
\nonumber \hat\rho_{\hat t}+\hat r (\hat \rho \hat V)_{\hat r}+3 \hat \rho \hat V&=0\,,\\
\nonumber \hat V_{\hat t} + \hat r \hat V \hat V_{\hat r}+\hat V^2 &=-\frac{3}{\hat r^3}\int_0^{\hat r}\hat \sigma^2(\hat \rho +\hat \Gamma(\hat \rho))\, d\hat \sigma\,,\\
\frac{dA}{d\hat t}&= \left[\frac{2}{A}+\frac{2\,\bar \alpha_0}{(\varpi+1) A^{\varpi+1}}\right]^{1/2}\,,
\end{align}
where the same symbol $A$ is used for the scale factor with respect to the original time and the new temporal variable $\hat t$. Here, the ordinary differential equation for the scale factor is coupled to the model because the nonlocal operator $\hat\Gamma$ depends upon $A$. 

As in Ref.~\cite{Chicone:2015sda}, we set $a_0=0$ so that $\tilde \alpha_0 \to  \alpha_0=2/(\lambda_0\, \mu_0)$ and the reciprocal kernel becomes
\begin{equation}\label{qker}
q(t, r):=\frac{\mu_0^2}{4 \pi \lambda_0 A^{\varpi}(t) } \frac{(1+\mu_0 r)e^{-\mu_0 r}}{(\mu_0 r)^2}\,.
\end{equation}
In dimensionless variables, using the same symbol $A$  for the scale factor in the new temporal variable, 
\begin{equation}\label{qkerdless}
\hat q(\hat t, \hat r):= \frac{\mu_0^2}  {4 \pi \lambda_0 A^\varpi (\hat t) }  \tilde q(\hat r)\,,
\end{equation}
where the new function $\tilde q$ is defined by
\begin{equation}\label{tq}
\tilde q (\hat r) = \frac{ (1+\hat r)e^{-\hat r} } {\hat r^2}\,.
\end{equation}

The operator $\Gamma$ in the new variables transforms to the dimensionless operator  $\hat \Gamma$ that appears in system~\eqref{sysdimless}.  In fact,  with $\hat \sigma=\mu_0\, \sigma$ 
\begin{align}\label{sysdimless2}
\nonumber \hat \Gamma(\hat \rho) (\hat t,\hat r)
&=  \frac{2 \pi} {\mu_0^3} 
       \int_0^\infty \Big (\int_{-1}^ 1 q(\tilde t_0 \hat t, \frac{ \sqrt{\hat r^2+\hat \sigma^2-2 \hat r\hat \sigma X}}  {\mu_0} )\,dX  \Big )\rho(\tilde t_0\hat t,\frac{ \hat \sigma}{\mu_0}) \hat \sigma^2\, d\hat \sigma\\
\nonumber  &= \frac{2 \pi}{\mu_0^3} 
      \int_0^\infty \Big (\int_{-1}^ 1 \hat q(\hat t, \sqrt{\hat r^2+\hat \sigma^2-2 \hat r\hat \sigma X})\,dX \Big  )\hat\rho(\hat t, \hat \sigma) \hat \sigma^2\, d\hat \sigma\\
&= \frac{1}{ 2 \lambda_0 \mu_0 A^\varpi (\hat t) }  
     \int_0^\infty  \Big (\int_{-1}^ 1 \tilde q(\sqrt{ \hat r^2+\hat \sigma^2-2 \hat r\hat \sigma X})\,dX \Big  )\hat\rho(\hat t, \hat \sigma)\hat \sigma^2\, d\hat \sigma\, .
      \end{align}
Alternatively,  the variables may be changed  in a similar manner  (perhaps most simply starting in the definition~\eqref{orgam} with $z=y-x$) to obtain 
\begin{align}\label{dimgam}
 \hat \Gamma(\hat \rho) (\hat t,\hat r)
\nonumber &= \frac{1}{ 2 \lambda_0 \mu_0 A^\varpi (\hat t) }  
     \int_0^\infty  
     \Big (
     \int_{-1}^ 1 (1+\hat{\beta}) e^{-\hat{\beta}}\hat\rho(\hat t, \sqrt{ \hat r^2+\hat{\beta}^2+2 \hat r  \hat{\beta} Y}) \,dY 
     \Big  )\, d \hat{\beta}\\
&=     \frac{\alpha_0}{ 4 A^\varpi (\hat t) }  
     \int_0^\infty  
     \Big (
     \int_{-1}^ 1 (1+\hat{\beta}) e^{-\hat{\beta}}\hat\rho(\hat t, \sqrt{ \hat r^2+\hat{\beta}^2+2 \hat r  \hat{\beta} Y}) \,dY 
     \Big  )\, d \hat{\beta}
\end{align}
where $\hat{\beta}$ and $Y$ are scalar dummy dimensionless variables. This formulation avoids division by zero at points where $\hat r^2+\hat \sigma^2-2 \hat r\hat \sigma X$ vanish. In particular the point where $\hat\sigma=\hat r$ and $X=1$ is problematic.  

The basic problem is to approximate the evolving density and velocity contrasts against the background solution of the system as defined in display~\eqref{F8} for an initial density contrast given by a Gaussian centered at $\hat r=0$ and zero initial velocity contrast with initial scaled time at an approximation of the decoupling era, which in this paper is taken to be  $A(\hat t)=10^{-3}$.

\subsection{The Numerical Code}

Because pressure does not appear in system~\eqref{sysdimless},  the local part of the PDE is not a strictly hyperbolic system of first order PDEs. Thus, the usual theory for systems of hyperbolic conservation laws does not apply directly.  In fact, as widely discussed in the literature (see, for example,~\cite{Bre})  the general PDEs for radially symmetric gas dynamics are not well understood.  On the other hand,  basic numerical methods usually used in computational fluid dynamics  are appropriate here---see, for background, in increasing order of sophistication~\cite{Chi, And, Wes}. 

Our numerical experiments are made using the finite difference approach.  Spatial partial derivatives are approximated with a convex combination of central differences and  upwinding  (using forward and backward differences according to the sign of the velocity) on a discretized finite spatial interval $0\le \hat {r}\le L$.  No artificial viscosity is added and flux limiters are not employed~\cite{LeV}.  Zero Neumann boundary conditions are imposed to respect the spherical symmetry at $\hat{r}=0$ via the usual method of computing only on interior nodes and setting the values of the state variables at the boundary node equal to their (computed) values at the adjacent interior node. 

System~\eqref{sysdimless} is posed on the entire half-line $\hat{r}\ge 0$.  Of course, numerical approximations are made on some finite domain.  The usual difficulties encountered in restricting a PDE model defined on an unbounded domain to a bounded domain for numerical work are compounded by the  presence of a nonlocal operation on density,  which in principle requires density data to be given at each moment of time on the entire half-line. The standard methods for restriction to a finite domain are radiation boundary conditions,  absorbing (sponge) layers, filtering and Dirichlet to Neumann boundary conditions~\cite{Giv}. None of these apply here in a straightforward manner. Nevertheless, a  stable numerical method can be  constructed based on physical considerations and the concept of an absorbing layer.  

At first glance, the presence of the negative exponential factor $e^{-\hat{\beta}}$ in the nonlocal operator~\eqref{dimgam} suggests that for sufficiently large $L$ the influence of this operator is negligible outside the computational domain  $0\le \hat{r}\le L$.  Indeed, the integration of $\hat{\beta}$ can be restricted to a finite interval $0\le \hat{\beta}\le L$ by accepting a controllably small error.   But, even with this restriction in force,  inspection of the quantity that appears under the square root in the argument of $\rho$ reveals that computation of the triple nonlocality integral  requires values of $\rho$ on $0\le \hat{r}\le 2 L$.  In fact,  to compute the truncated  integral exactly,  values of the density on the subdomain $\{(\hat r,\hat{\beta}, Y) : \hat r^2+\hat{\beta}^2+2 \hat r  \hat{\beta} Y \ge L^2\}$ of the  rectangular domain of the triple integration are required. The size of this domain remains proportionally the same with increasing $L$.

The desired solution is initiated by a perturbation localized near $\hat r=0$ of the 
space-independent background solution. The influence of this perturbation does not spread with infinite speed. But as mentioned,  the nonlocality does take into account values of the density outside every finite domain. To mitigate the error inevitably produced by using a finite computational domain, we use an auxiliary function of position to smoothly do away with the terms in the model system  near the right-hand boundary of the computational domain so that the system is smoothly transformed to the corresponding system of ODEs that has the  background as an exact solution. Values of the density within the computational domain are those produced by the time evolution of the modified system and those  outside the computational domain are taken to be the corresponding values of the background.  For this scenario, Dirichlet boundary conditions are used for the modified system  at $\hat r=L$ where  the computed state variables are required to match their background values.  Our choice for the auxiliary function, which simply multiplies each term that contains a spatial derivative, has unit value over the subinterval $0\le \hat{r}\le 0.95\, L$  and dies off to zero on the remaining $5\%$ of the computational domain via a continuously differentiable cubic polynomial extension.  The use of this transitional (sponge) layer produces stable numerics and physically reasonable results for the evolving state variables. 

The nonlocal term is approximated as an iterated triple integral using Simpson's rules. Values of $\hat \rho$ required to evaluate the integrand of the inner integral at non-grid points along the spatial interval, see Eq.~\eqref{dimgam},  are obtained from  cubic spline interpolations computed using the current values of $\hat \rho$ at the nodes.  Fewer nodes than in the complete spatial discretization are used to create the cubic spline to avoid possible numerical roundoff error associated with solving the large tridiagonal linear system required to obtain the splines.  Clamped splines  (see, for example,~\cite{BF}) are used with boundary values that are compatible with the spatial boundary conditions. In particular,  the clamped spline has zero derivative at $\hat r=0$ and the right-hand derivative is approximated using a three-point numerical differentiation formula.  

Modified (i.e. improved) Euler  time stepping  is employed. The time-stepping algorithm uses a simple CFL condition $\Delta t<0.5\, \Delta \hat{r}/\abs{x}$, where $\abs{x}$ is the maximum norm of the current state $x$,  to determine a reasonable step size. Due to the large size of the background 
density at time zero of integration (i.e.\ background $\hat{\rho}$ of usually about $10^9$ at  the decoupling epoch),
   a very small starting step size is required. After a preassigned number of steps, the CFL condition is again checked. When it suggests a larger step size  can be used to maintain stability,  the step size $\Delta t$ is ramped up to this new size $ 0.5\, \Delta \hat{r}/ \abs{x}$ over the steps preceding the next CFL check. This conservative method  for step size control generally produces smooth results.  In a typical run, the starting (time) step size is of order  $10^ {-12}$ and the final step size  is $10^{-4}$. 
  
\subsection{Results of Numerical Experiments}

The main system parameters  are $\delta$ and $\hat w$ of the initial Gaussian density contrast---see Eqs.~\eqref{F19} and~\eqref{F20}, the memory fade exponent $\varpi$ and the choice of the maximum spatial distance $L$.  For the numerical algorithm, the relevant parameters are the number of grid points $m+1$ on the interval $[0,L]$, the number of grid points $n$ on the interval from $[-1,1]$ for discretization of the inner integral in the computation of $\hat \Gamma$, the weight parameter for central differences versus upwind differences,  the number of grid points used for spline interpolation, the number of steps between step-size changes, the CFL step-size change rule, the function determining the grid points where pure upwind differences are used,  the function that determines the sponge layer and the boundary conditions. In Figures 4 and 5, the standard data is defined to be 
\begin{gather}
 A(0)=10^{-3}\,, \qquad  \delta=10^{-5}\,,    \qquad        \alpha_0 = 11\,.
\end{gather}

A comment is in order here regarding the fact that shocks are prevalent in the numerical treatment of gas dynamics. The numerical methods employed here are known to be reasonable as  shock capturing techniques. But, perhaps due to taking initial data near the background Hubble flow, shocks do not seem to appear in the numerical experiments reported here.    

\subsubsection{Change in Width of Gaussian}

Figure 4 shows the results of a numerical experiment with standard data together  with  $\varpi=0.5$  and several choices for the width of the initial Gaussian density contrast.  The  length of the spatial domain is $L=40$. As the width of the Gaussian grows, the density contrast at the present epoch defined by the scale parameter $A=1$ also grows and approaches the terminal value corresponding to a uniform initial density contrast over $L$. Let us note in this connection that  the theoretical density contrast on the infinite interval in the linear approximation for the corresponding spatially independent growing-mode solution given by Eq.~\eqref{F27}  with $A=1$ is given by $\mathbb{D}(1) \approx 0.128$.   

\begin{figure}
\begin{center}
\includegraphics[height= 6 cm, width=12 cm]{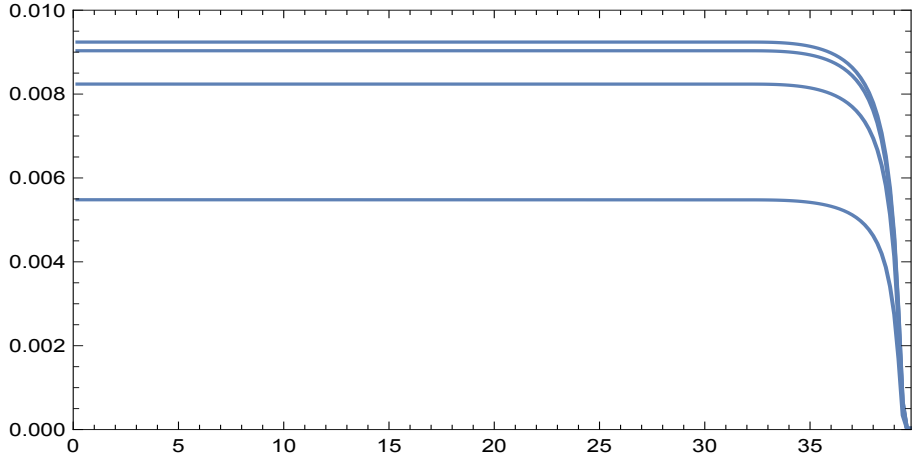}
\end{center}
\caption{The graphs are density contrast versus spatial position $\hat r= \mu_0\,r$, $0 \le \hat r \le 40$, for the dust model at the present epoch with the scale parameter $A=1$. We assume the standard data together with $\varpi=0.5$. The initial density contrast is given by the Gaussian in Eqs.~\eqref{F19} and~\eqref{F20} with amplitude $\delta = 10^{-5}$ and various values of the width parameter $\hat w$. The depicted graphs are for the width parameter values 4, 8, 16 and 32 from bottom to top. }
\label{fig4} 
\end{figure}

\subsubsection{Change in Spatial Cutoff}

\begin{figure}
\begin{center}
\includegraphics[width=12 cm]{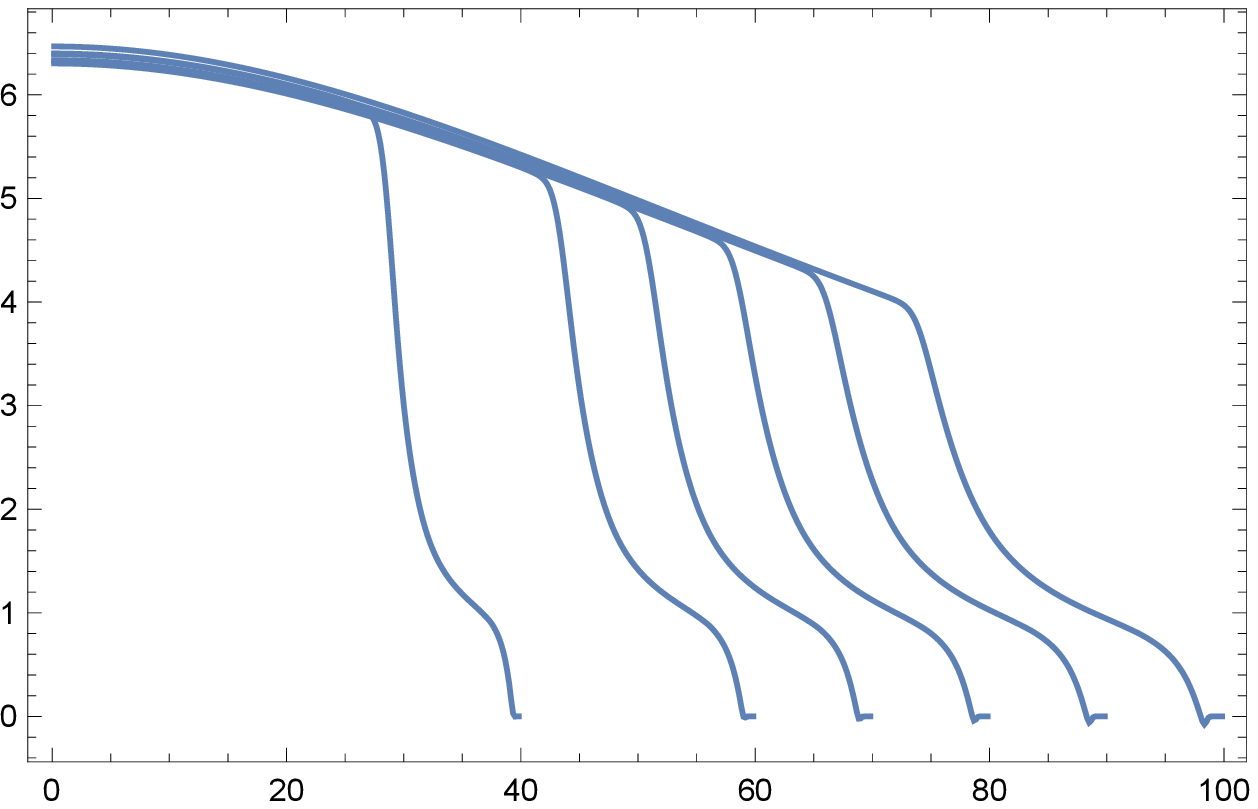}
\end{center}
\caption{ The graphs are density contrast versus spatial position $\hat r= \mu_0\,r$ for the dust model    at an epoch with the scale parameter $A=0.475$. The initial density contrast is given by the Gaussian in Eqs.~\eqref{F19} and~\eqref{F20} with amplitude $\delta =10^{-5}$ and width $\hat w=1$.  We assume the standard data together with $\varpi=2.0$. From bottom to top, the graphs are computed on computational domains $0\le \hat r\le L$  for $L=40,60,70,80,90,100$.  }
\label{fig5} 
\end{figure}

Figure 5 shows the density contrast approximated by  numerical experiments with standard data together  with  $\varpi=2.0$  and several choices for the length of the discretized spatial interval  given by $L=40,60,70,80,90,100$. The terminal scale factor was set at  $A=0.475$, i.e.\ long before the present era,   to show a terminal density contrast that exceeds five and hence possibly leads to structure formation~\cite{Mu}. Due to the finite spatial domain and the sponge layer treatment at the right-hand boundary,  each graph has a fictitious portion near the corresponding position $\hat r=L$. On the other hand, convergence to an approximation of the corresponding solution of the model equation as $L$ increases seems to occur.

\subsection{Interpretation of Numerical Results}

The nonlocal Poisson equation that we have employed in our nonlocal Euler--Poisson toy model is physically equivalent to the gravitational force law~\eqref{DI1}. This attractive interaction between two point masses has the characteristic feature that it modifies the Newtonian inverse square force law over galactic scales; however, for $r \gg \mu_0^{-1} \approx 17$ kpc, it approaches the inverse square law but with a modified Newtonian gravitational ``constant" given by
\begin{equation}
\tilde{G}(t) = G\,[1+ \tilde{\alpha}_0\,A^{-\varpi}(t)]\,,
\end{equation}
as explained in detail in Section V. The import of this circumstance for nonlocal Newtonian cosmology in connection with Newton's shell theorem is that the force of nonlocal gravity inside a spherically symmetric cavity cannot in general be ignored. This point is illustrated via a simple example in Appendix A. 

It follows from Newton's shell theorem that  in \emph{local} Newtonian cosmology, the homogeneous and isotropic universe exerts no gravitational force on a spherically symmetric over-dense region around $r=0$ that separates from the background.  The resulting structure is then expected to collapse under its own gravity. On the other hand, in \emph{nonlocal} Newtonian cosmology, the internal gravitational attraction of the over-dense region is somewhat offset by the external gravitational attraction of the cosmological background inside a spherical shell of thickness 
$\sim \mu_0^{-1}$ that immediately surrounds the over-dense region.  While the inner part of the over-dense region experiences gravitational instability in connection with the growing mode, its outer part can be significantly affected by the external attraction. As the over-dense region is stretched out in this way, its outer parts are being continually attracted by the cosmological background in outer shells each with thickness  of order $\mu_0^{-1}$ and ever larger radii. The outer boundary of such shells should extend all the way out to $\hat r = \infty$.  This heuristic picture 
appears to provide a reasonable physical interpretation of our numerical results presented in Fig. 5.

\begin{acknowledgments}

BM is grateful to Roy Maartens and Haojing Yan for valuable discussions.  

\end{acknowledgments}

\appendix{}

\section{Nonlocal Gravity Violates Newton's Shell Theorem}\label{appA}

Imagine a test particle $\mathbb{P}$ of unit mass at rest inside a spherically symmetric cavity of radius $\mathbb{R}_0$ at the present epoch. A thin shell of negligible thickness and total mass $\mathbb{M}$ uniformly surrounds the cavity. According to Newton's shell theorem, there is no force of gravity inside the cavity; therefore, 
$\mathbb{P}$ is expected to remain at rest. On the other hand, according to nonlocal gravity theory in the Newtonian regime, there is a force of gravity on $\mathbb{P}$ that can be calculated straightforwardly using Eq.~\eqref{DI1} \emph{at the present epoch}. To this end, let us choose the Cartesian coordinate system such that its origin is at the center of the cavity and $\mathbb{P}$ is initially at rest on the positive $z$ axis. Moreover, we let $\zeta := z/\mathbb{R}_0$ and we measure time $t$ in units of $(\mathbb{R}_0^3/G\,\mathbb{M})^{1/2}$. The Newtonian equation of motion for $\mathbb{P}$ can then be expressed as
\begin{equation}\label{A1}
\frac{d^2\zeta}{dt^2} = \frac{\alpha_0}{4\,\zeta^2}\,[\mathbb{W}_1 + \mathbb{W}_2]\,,
\end{equation}
where $\alpha_0 = 11$ and $1>\zeta>-1$.  In Eq.~\eqref{A1}, $\mathbb{W}_1$ is given by
\begin{equation}\label{A2}
\mathbb{W}_1 = \int_{1-\zeta}^{1+\zeta} \left(1-\frac{1-\zeta^2}{x^2}\right)(1+\frac{1}{2}\,r_0\,x)\,e^{-r_0\,x}\,dx\,
\end{equation}
where 
\begin{equation}\label{A3}
r_0 := \mu_0\,\mathbb{R}_0\,. 
\end{equation}
In fact, $\mathbb{W}_1$ can be expressed as
\begin{equation}\label{A4}
\mathbb{W}_1 =- [3 \zeta \cosh(r_0\,\zeta) + \left(1-\frac{3}{r_0}\right) \sinh(r_0\,\zeta)]\,e^{-r_0} +\frac{1}{2} r_0 (1-\zeta^2) [E_1(r_0 - r_0\,\zeta) -  E_1(r_0 + r_0\,\zeta)]\,.
\end{equation}
Moreover, $\mathbb{W}_2$ is given by
\begin{equation}\label{A5}
\mathbb{W}_2 =  \int_{1-\zeta}^{1+\zeta} \left(1-\frac{1-\zeta^2}{x^2}\right)\,\mathbb{U} (x)\,dx\,,
\end{equation}
where $\mathbb{U} = \mathcal{E}/\alpha_0$. We recall from the discussion in Section V that $\mathcal{E}$ is given by either $\mathcal{E}_1$ or $\mathcal{E}_2$ depending on whether we employ reciprocal kernel $q_1$ or $q_2$. Thus,  
\begin{equation}\label{A6}
\mathbb{U}_1(x) = \frac{1}{2}\,\varsigma\, e^{\varsigma}\Big[E_1(\varsigma)-E_1(\varsigma+r_0\, x)\Big]\,, \qquad \mathbb{U}_2 (x) = 2\, \mathbb{U}_1(x) - \frac{1}{2}\,\varsigma\,\frac{r_0\, x}{\varsigma+r_0\,x}e^{-r_0\,x}\,,
\end{equation}
where we assume that $\varsigma = 10^{-4}$.

\begin{figure}[h]
\begin{center}
\includegraphics[width=12 cm]{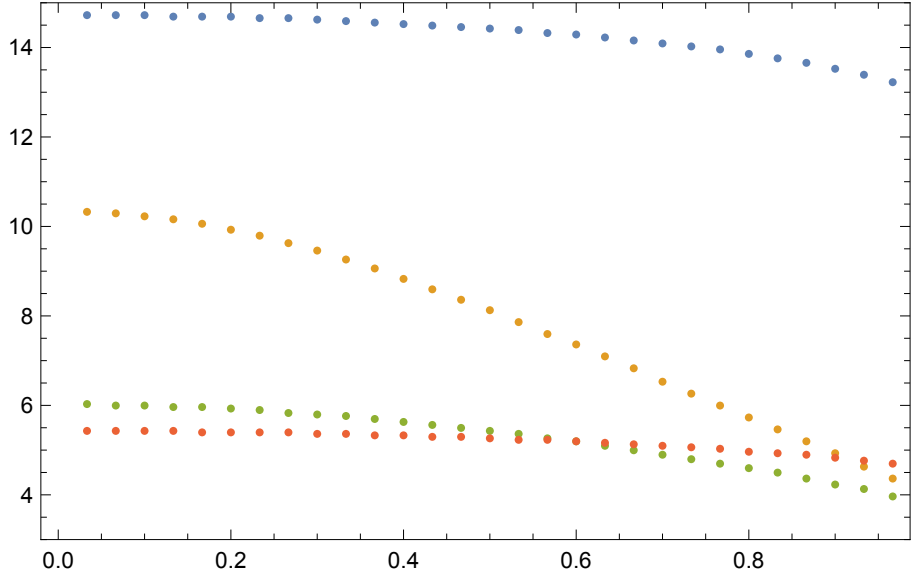}
\end{center}
\caption{The period of the motion is plotted here versus $\zeta_0$, $0<\zeta_0<1$, for different values of $r_0$. Starting near $\zeta_0 = 0$, from bottom to top, the graphs represent $r_0 = 1, 3, 5, 0.1.$}
\label{fig6} 
\end{figure} 

Let us first choose $\mathbb{U}_1(x)$ for the specific calculations in this appendix.  The singularity at $\zeta=0$ in Eq.~\eqref{A1} is removable; in fact, near $\zeta=0$ we have
\begin{align}\label{Aforce} 
\nonumber \frac{\alpha_0}{4\,\zeta^2}\,[\mathbb{W}_1 + \mathbb{W}_2]=
{}&-\frac{\alpha_0 r_0^2 (1+r_0+\varsigma) e^{-r_0}} {6(r_0+\varsigma)}\,\zeta
 -\frac{\alpha_0 r_0 e^{-r_0} }{60 (r_0+\varsigma)^3}\, p_{\varsigma}(r_0)\zeta^3\\ 
 &+O(\zeta^5)\,,
\end{align}
where
\begin{align}\label{A8} 
\nonumber p_{\varsigma}(r_0)= {}& r_0^6 + 3 r_0^5 (1+\varsigma) +r_0^4(4+8\varsigma +3 \varsigma^2)+r_0^3(4+10\varsigma+7\varsigma^2+\varsigma^3) \\
 &+2 r_0^2\varsigma(5+5 \varsigma+\varsigma^2)+4 r_0\varsigma^2(3+\varsigma)+4 \varsigma^3\,.
\end{align}

The Newtonian equation of motion~\eqref{A1} has a first integral. Moreover the  symmetry of the configuration, reflected in the fact that the right-hand side of the differential equation is an odd function of $\zeta$,  implies that orbits in the phase plane are symmetric about the coordinate axes. These facts and the form of the series representation~\eqref{Aforce} imply that this dynamical system has a  center at the origin in the phase plane surrounded by
 an annulus of periodic orbits~\cite{CH}. This period annulus appears to contain all orbits  of test particles  starting  at $t=0$ with  $\zeta = \zeta_0$, where $\zeta_0 > 0$ is a constant, and $d\zeta/dt = 0$. Thus, every such test particle moves periodically from $\zeta_0$ to $- \zeta_0$ and back forever.  For $\zeta_0 = 0$, however, $\zeta=\zeta_0=0$, and the particle remains stationary at the geometric center of the cavity surrounded by the  material spherical shell. 

Using the series representation~\eqref{Aforce}, the limiting period of periodic orbits as they approach the center is given by 
\begin{equation}\label{Alimper}
\frac{2\,\pi}{r_0}\,\sqrt{\frac{6\,(\varsigma+r_0)}{\alpha_0\,(1+\varsigma+r_0)}}\,e^{r_0/2}\,.
\end{equation} 
Also, using~\cite[Lemma 4.1]{CHJA} and the series representation~\eqref{Aforce}, a lengthy but straightforward computation (perhaps aided by the use of a computer algebra system) can be used to show that the period function decreases as a function of position along the positive $\zeta$ axis in the phase plane near $\zeta=0$.

Extensive numerical work suggests that for a cavity with fixed radius $r_0$, the period of the oscillation \emph{decreases} (globally) with
 increasing $\zeta_0$. This circumstance is illustrated in Figure 6.  These numerical computations agree with the exact limiting period given by expression~\eqref{Alimper}. It is possible that the global behavior of the period function illustrated in Fig. 6 can be proved analytically; however, this is a task that is beyond the scope of the present work~\cite{CHC, CHJA}.
 
If we use $\mathbb{U}_2(x)$ instead of  $\mathbb{U}_1(x)$, everything turns out  to be
\emph{qualitatively} the same as before. For example, the limiting period at the center is now
\begin{equation}\label{Alimper2}
\frac{2\,\pi}{r_0}\,\sqrt{\frac{6\,(\varsigma+r_0)^2}{\alpha_0\,r_0\,(1+\varsigma+r_0)}}\,e^{r_0/2}\,.
\end{equation} 
The period function near the center again decreases near this point as $\zeta_0$ increases.  There is certainly no significant difference between the behavior of the two reciprocal kernels that shows up in these computations. 

Finally, we recall from Section II that the parameter $\varsigma$ is such that 
$0< \varsigma< 2/\alpha_0$. It appears that the validity of our main numerical results extends beyond $\varsigma=10^{-4}$. In fact,  preliminary numerical work indicates that for fixed $\varsigma$ and $r_0$, the period function monotonically decreases with increasing $\zeta_0$.

\section{Distention of the Kepler System with slowly Decreasing Mass}\label{appB}

Consider the two-body problem with a time-dependent gravitational potential $\Phi$. The bound planar orbit can be described in terms of the relative polar coordinates $r$ and $\phi$ such that $r^2\,\dot \phi=\mathbb{L}$, where $\mathbb{L}$ is the constant specific orbital angular momentum of the two-body system. The remaining radial equations of motion are given by
\begin{align}\label{B1}
\nonumber \dot r&=v, \\
 \dot v&= \frac{\mathbb{L}^2}{r^3}-\frac{\partial \Phi}{\partial r} (r, A),
\end{align}
where $\dot A$ is given by Eq.~\eqref{N8} with $\bar{E} = 0$. Let us suppose for the moment that $A$ is a parameter that does not depend upon time.  Then system~\eqref{B1} has a unique center in the phase plane corresponding to the circular orbit  at  $(r_c, 0)$,   the point where the vector field vanishes. Of course, the position of $r_c$ depends on the choice of $A$.

We are interested in the gravitational inverse square law; therefore, in our Kepler system  
\begin{equation}\label{B2}
 \frac{\partial \Phi}{\partial r} = \frac{\gamma_0 \,\tilde{\beta}^\varpi}{r^2}\,, \qquad \tilde{\beta}:= \frac{1}{A(t)}\,.
\end{equation}
Moreover, we assume that at some initial cosmic epoch $t=t_{in}$, the relative orbit is an ellipse with semimajor axis $\mathbb{A}_{in}$, eccentricity $e$ and Keplerian frequency $\omega_{in}$ such that 
\begin{equation}\label{B3}
r = \frac{\mathbb{A}_{in}\,(1-e^2)}{1+e\,\cos \phi}\,, \qquad \mathbb{L}=\mathbb{A}_{in}^2\,\omega_{in}\,(1-e^2)^{1/2}\,.
\end{equation}
To proceed, it is useful to specify initial conditions for our dynamical system as follows:
\begin{equation}\label{B4}
t=t_{in}\,,\qquad \phi=0\,, \qquad r = \mathbb{A}_{in}\,(1-e)\,, \qquad v =0\,, \qquad 
\mathbb{A}_{in}^3\,\omega_{in}^2 =\frac{\gamma_0}{A^\varpi(t_{in})}\,.
\end{equation}
These initial conditions correspond to motion in the positive sense starting from the pericenter of the osculating ellipse at $t=t_{in}$. 

It proves useful to introduce dimensionless quantities
\begin{equation}\label{B5}
\tilde{t}=\frac{\omega_{in}\,(t-t_{in})}{2\,\pi}\,, \qquad \tilde{r} = \frac{r}{\mathbb{A}_{in}}\,, \qquad \tilde{v}=\frac{2\,\pi\,v}{\omega_{in}\,\mathbb{A}_{in}}\,, \qquad 
\epsilon=\frac{2\,\pi}{\omega_{in}}\,\left(\frac{8\,\pi\,G\,\rho_0}{3}\right)^{1/2}\,,
\end{equation}
where $\epsilon$, $0<\epsilon\ll 1$ is the ratio of the ``fast" period of the initial Keplerian ellipse to the ``slow" Hubble period characteristic of the background cosmological model---cf. Eq.~\eqref{N9}.  Our system of equations can now be expressed as
\begin{align}\label{B6}
\nonumber \frac{d\tilde{r}}{d\tilde{t}}&=\tilde{v}\,, \\
\nonumber \frac{d\tilde{v}}{d\tilde{t}}&=\frac{4\,\pi^2\,(1-e^2)}{\tilde{r}^3}-\frac{h\,\tilde{\beta}^\varpi}{\tilde{r}^2}\,,\\
\frac{d \tilde{\beta}}{d\tilde{t}}&=-\epsilon\,\mathbb{G}(\tilde{\beta})\,,
\end{align}
where $h$ is a positive constant given by
\begin{equation}\label{B7}
h= 4\,\pi^2\, A^\varpi(t_{in})\,
\end{equation}
and $\mathbb{G}$ can be expressed as
\begin{equation}\label{B8}
\mathbb{G}(\tilde{\beta})= \tilde{\beta}^2 \,\left(\tilde{\beta} +\frac{\tilde{\alpha}_0}{\varpi + 1}\,\, \tilde{\beta}^{\varpi+1}\right)^{1/2}\,.
\end{equation}
In this setting $\tilde{\beta}$ decreases with the temporal parameter and thus the system may be viewed as the Keplerian two-body problem with decreasing mass.
The initial data is now
\begin{equation}\label{B9}
\tilde{t}=0\,, \qquad \tilde{r}(0)=1-e\,, \qquad \tilde{v}(0)=0\,, \qquad \tilde{\beta}(0)=\frac{1}{A(t_{in})}>1\,.
\end{equation}
The mathematical problem is to find the osculating ellipse at the present epoch $t=t_0$ when $ \tilde{\beta}=1$ and thereby compute the semimajor axis of this instantaneous Keplerian ellipse in order to determine the distention of the initial instantaneous Keplerian ellipse in cosmic time: $t_{in} \to t_0$.

Following the standard prescription~\cite{SVM},  we now use variables $(\check{u}, \phi)$ defined by
\begin{equation}\label{B10} 
\check{u}=\frac{1}{\tilde{r}}, \qquad  \frac{d\phi}{d \tilde{t}}=2\,\pi\,(1-e^2)^{1/2}\, \check{u}^2\,,
 \end{equation}
with $\phi(0)=0$ and  
\begin{equation}\label{B10a} 
 \tilde{v} = - 2\,\pi\,(1-e^2)^{1/2}\, \check{v}\,,
 \end{equation}
to rewrite the model equations in the perturbed harmonic oscillator form, namely, 
\begin{align}\label{B11}
\nonumber \frac{d\check{u}}{d\phi}&=\check{v}\,,\\
\nonumber \frac{d \check{v}}{d\phi}&=-\check{u} + C\,, \\
 \frac {d \tilde{\beta}}{d\phi} &=-\epsilon \frac{\mathbb{G}(\tilde{\beta})}{2\,\pi\,(1-e^2)^{1/2}\, \check{u}^2}\,,
\end{align}
where $C$ is given by
\begin{equation}\label{B12} 
 C= \frac{A^\varpi(t_{in})}{1-e^2}\,\tilde{\beta}^\varpi\,.  
 \end{equation}

Next, the van der Pol transformation~\cite{SVM} to the new variables $(P, Q)$,
\begin{equation}\label{B13}
 \check{u}= C+P \cos \phi +Q \sin\phi\,, \qquad \check{v}=-P \sin \phi +Q \cos \phi\,,
 \end{equation}
 recasts  system~\eqref{B11} into a form suitable for averaging:
\begin{align}\label{B14}
\nonumber \frac{dP}{d\phi}&=\epsilon \frac{A^\varpi(t_{in})}{2\,\pi\,(1-e^2)^{3/2}}\, \frac{\varpi \tilde{\beta}^{\varpi-1} \mathbb{G}(\tilde{\beta}) \cos \phi}{(C +P \cos \phi+Q \sin\phi)^2}\,,\\
\nonumber \frac{dQ}{d\phi}&=\epsilon \frac{A^\varpi(t_{in})}{2\,\pi\,(1-e^2)^{3/2}}\,\frac{\varpi \tilde{\beta}^{\varpi-1} \mathbb{G}(\tilde{\beta}) \sin \phi}{(C +P \cos \phi+Q \sin\phi)^2}\,,\\
\frac{d\tilde{\beta}}{d\phi} &=-\epsilon \frac{1}{2\,\pi\,(1-e^2)^{1/2}}\, \frac{\mathbb{G}(\tilde{\beta})}{(C +P \cos \phi+Q \sin\phi)^2}\,.
\end{align}
Using the fact that 
\begin{equation}\label{B15} 
 \frac{1}{2\,\pi} \int_0^{2\,\pi} \frac{dx}{1+\tau \cos(x+\theta)}= \frac{1}{(1-\tau^2)^{1/2}}\,, \qquad |\tau| <1\,,  
\end{equation}
and averaging over the fast angle $\phi$ produces the first-order averaged system
\begin{align}\label{B16}
\nonumber \frac{d\mathcal{P}}{d\phi}&= -\epsilon \frac{A^\varpi(t_{in})}{2\,\pi\,(1-e^2)^{3/2}}\, \frac{\varpi \beta^{\varpi-1} \mathbb{G}(\beta) \mathcal{P}}{(C^2 -\mathcal{P}^2- \mathcal{Q}^2)^{3/2}}\,,\\
\nonumber \frac{d\mathcal{Q}}{d\phi}&= -\epsilon \frac{A^\varpi(t_{in})}{2\,\pi\,(1-e^2)^{3/2}}\,\frac{\varpi \beta^{\varpi-1} \mathbb{G}(\beta) \mathcal{Q}}{(C^2 -\mathcal{P}^2- \mathcal{Q}^2)^{3/2}}\,,\\
\frac{d \beta}{d\phi} &=-\epsilon \frac{A^\varpi(t_{in})}{2\,\pi\,(1-e^2)^{3/2}}\, \frac{\beta^\varpi \,\mathbb{G}(\beta)}{(C^2 -\mathcal{P}^2- \mathcal{Q}^2)^{3/2}}\,.
\end{align}
Here, 
\begin{equation}\label{B17} 
\mathcal{P} =\,  < P >\,,  \qquad \mathcal{Q} =\, < Q >\,,  \qquad \beta = \,< \tilde{\beta} >\,  
\end{equation}
and the initial data are given by
\begin{equation}\label{B18} 
\mathcal{P}(0) = P(0) = \frac{e}{1-e^2}\,,  \qquad \mathcal{Q}(0) = Q(0)=0\,,  \qquad \beta(0) =  \tilde{\beta}(0) = \frac{1}{A(t_{in})} >1\,.  
\end{equation}
By the averaging theory, 
\begin{equation}\label{B19}
P(\phi)= \mathcal{P}(\phi)+O(\epsilon)\,, \qquad Q(\phi)= \mathcal{Q}(\phi)+O(\epsilon)\,, \qquad  \tilde{\beta}(\phi)= \beta(\phi)+O(\epsilon)\,,
\end{equation}
on an integration scale of order $1/\epsilon$; that is, for some constant $T$ (independent of $\epsilon$) the estimate is valid for sufficiently small $\epsilon$ as long as $0\le \epsilon \phi \le T$. 

Although the averaged system~\eqref{B16} is nonlinear, it has some special properties that can be exploited; for instance, it follows from the averaged system that 
\begin{equation}\label{B20} 
\mathcal{Q} \frac{d\mathcal{P}}{d\phi}=\mathcal{P}  \frac{d\mathcal{Q}}{d\phi}\,;
\end{equation}  
hence, $\mathcal{Q}/\mathcal{P}$ is constant. Therefore,  for the case at hand, where $\mathcal{Q}$ is initially zero but $\mathcal{P}$ is not,  $\mathcal{Q}$ is zero as long as it is defined.  

For the simplest case, where the Keplerian orbit is initially a circle ($e=0$),  both $\mathcal{P}$ and $\mathcal{Q}$ vanish as the averaged system evolves. This choice of initial data occurs at the minimum of the effective potential energy.  An exact formula for the dependent variable $\beta$  remains implicitly defined as the  nonzero solution of a nonlinear initial value problem; but, for the present case, the relevant value of interest is $\beta=1$  at the present epoch. Because $\mathcal{P}$ and $\mathcal{Q}$ remain equal to zero, averaging produces the approximations
$P(\phi)=O(\epsilon)$ and  $Q(\phi)=O(\epsilon)$ on the angular $\phi$ scale of $1/\epsilon$. This means that the orbit remains essentially a circle and the approximate value of $\tilde{r}=r / \mathbb{A}_{in}$ (up to an error of order $\epsilon$) at the stopping point of the evolution is $1/\check{u}= 1/C=1/A^\varpi(t_{in})$, see Eq.~\eqref{B12}. 
Thus $r$ increases (approximately) from an initial value of $\mathbb{A}_{in}$ to a final value of 
$\mathbb{A}_{in}/A^\varpi(t_{in})$ over the specified interval of time ($t_{in} \to t_0$). 

For the general case, where we start from a Keplerian ellipse, the main observation akin to the proportionality of $\mathcal{P}$ and $\mathcal{Q}$ is the relation between
$\mathcal{P}$ and $\beta$. By inspection,
\begin{equation}\label{B21} 
\beta \frac{d\mathcal{P}}{d\phi}=\varpi \mathcal{P}  \frac{d\beta}{d\phi}\,. 
\end{equation}
Thus,  $\mathcal{P}/\beta^\varpi$ is constant.  In view of the initial data, 
\begin{equation}\label{B22} 
\mathcal{P}(\phi)=\frac{e\,A^\varpi(t_{in})}{1-e^2}\,\beta^\varpi(\phi)\,.
\end{equation}
It then follows from Eq.~\eqref{B13} that
\begin{equation}\label{B23} 
\check{u} =\frac{A^\varpi(t_{in})}{A^\varpi(t)}\,\frac{1+e\,\cos \phi}{1-e^2}\,,
\end{equation}
which means, when compared to the initial orbit in Eq.~\eqref{B3},  that as the universe expands, the shape of the orbit remains essentially the same, but  its dimensions  increase in proportion to 
\begin{equation}\label{B24} 
\frac{A^\varpi(t)}{A^\varpi(t_{in})}\,.
\end{equation}
Moreover, this general result is independent of the specific formula for the Hubble expansion of the universe that depends upon $\mathbb{G}(\beta)$. At the present epoch $t=t_0$,  $A(t_0)=1$ and thus the orbit is  approximately an ellipse with eccentricity $e$ and semimajor axis $\mathbb{A}_{in}/A^\varpi(t_{in})$. Therefore,  the distention of the elliptical orbit over the time interval $t_{in} \le t \le t_0$ varies in inverse proportion to the attractive force of gravity, see Eq.~\eqref{B2}. 

\section{A Solvable Toy Model}\label{appC}

Consider a special toy model given by a variant of system~\eqref{rdmrecast} where the nonlocal part is replaced by $k^2(t)$ that represents a certain attractive interaction. The model under consideration here is thus given by 
\begin{align}\label{C1}
\nonumber
\rho_t+ r (\rho V)_r+ 3 \rho V&=0\,,\\
V_t+r VV_r+V^2 &=- k^2(t)\,.
\end{align}
We seek a solution that is analogous to the Hubble flow; therefore, we assume $V=V(t)$. Hence, the radial fluid velocity is given by $v=r\,V(t)$, where $V$ is the analog of the Hubble parameter in this case. 

Let us  choose a  function $f(t)$ such that for $t>0$,
\begin{equation}\label{C2}
f(t) > 0\,, \qquad f(0)=1\,, \qquad \dot{f}(0) = V_0 >0\,, \qquad \ddot{f}(t)<0 \,.
\end{equation}
Furthermore, we assume 
\begin{equation}\label{C3}
V(t) = \frac{\dot f}{f}\,,
\end{equation}
so that  $f(t)$ is the analog of the cosmological scale factor and the Euler equation then reduces to 
\begin{equation}\label{C4}
\ddot f +k^2(t)\, f =0\,.
\end{equation}
Inserting $V=\dot{f}/f$ into the continuity equation  and applying  the method of characteristics (see, for example, Ref.~\cite{E}), we obtain the density of the fluid
\begin{equation}\label{C5}
\rho(t, r) = \frac{1}{f^3(t)}\, \tilde{\rho}\left(\frac{r}{f(t)}\right)\,,
\end{equation}
where  $\tilde{\rho}(r)$ is the initial density at $t=0$, i.e. $\rho(0, r) :=\tilde{\rho}(r)$.  

For a simple explicit solution, let us assume that $k$ is a positive constant; then,  Eq.~\eqref{C4} implies
\begin{equation}\label{C6}
f(t) = \cos(k\,t) +\frac{V_0}{k}\,\sin(k\,t)\,.
\end{equation}
Moreover, let the initial density be the linear superposition of a uniform background of density $c_0>0$ and a Gaussian perturbation of constant amplitude $\varepsilon >0$, namely, 
\begin{equation}\label{C7}
\tilde{\rho}(r)  =  c_0 + \varepsilon\, \exp{\left(-\frac{r^2}{2\,b_0^2}\right)}\,,
\end{equation}
where $b_0>0$ is another constant. 
Then, it follows from Eq.~\eqref{C5} that 
\begin{equation}\label{C8}
\rho(t, r)= \frac{c_0}{f^3(t)} +\rho_{\varepsilon}(t, r)=\frac{c_0}{f^3(t)} + \frac{\varepsilon}{f^3(t)}\, \exp{\left(-\frac{r^2}{2\,b_0^2\,f^2(t)}\right)}\,,
\end{equation}
where $f(t)$ is given by Eq.~\eqref{C6}. That is, for $t>0$ the matter density consists of the background $c_0/f^3(t)$ and the Gaussian perturbation $\rho_{\varepsilon}(t, r)$ given by Eq.~\eqref{C8}. 

It is interesting to compare and contrast this local toy model with our nonlocal cosmological model.  For instance, unlike the cosmological case, the background and the perturbation here evolve with time independently of each other. As time increases, $f(t)$ eventually goes to zero when $t\to t_c$,  
\begin{equation}\label{C9}
t_c =  \frac{\pi}{2\,k} + \frac{1}{k}\, \arctan{ \left(\frac{V_0}{k}\right)}\,.
\end{equation}
Under the influence of the attractive interaction,
the background density eventually becomes unbounded everywhere but the Gaussian perturbation simply collapses to a Dirac delta function singularity  at the origin.
 To see this, note that as $t \to t_c$, $\rho_{\varepsilon}(t, r>0) \to 0$ and $\rho_{\varepsilon}(t, 0) \to \infty$, while  the total mass of the perturbation $M_{\varepsilon}$ is fixed for all time, namely,
\begin{equation}\label{C10}
M_{\varepsilon} = 4\,\pi \int_0^\infty \rho_{\varepsilon}(t, r)\,r^2\,dr  = (2\,\pi)^{3/2}\, b_0^3\,\varepsilon\,.
\end{equation}

\end{document}